\documentclass[review]{elsarticle}
\usepackage{changes}
\usepackage{pifont}
\usepackage{verbatim}
\usepackage{lineno,hyperref}
\usepackage{float}
\usepackage{graphicx}
\usepackage{subfigure}
\usepackage{graphicx}
\usepackage{geometry}
\usepackage{amsfonts}
\usepackage{indentfirst}
\usepackage{bm}
\usepackage{amsmath}
\usepackage{amssymb}

\usepackage[ruled,linesnumbered]{algorithm2e}

\usepackage{multirow} 
\usepackage{color}
\usepackage{enumerate}

\usepackage{graphicx}
\usepackage{float}
\modulolinenumbers[5]

\date{September 24, 2021}
\journal{arXiv}

\begin{document}

\begin{frontmatter}

\title{A minimalistic approach to simulate multiple failure mechanisms of metal matrix fiber-reinforced composites by the phase field and cohesive zone models }

\author{Zhaoyang Hu\textsuperscript{1}\footnote{\textsuperscript{1}These authors contributed equally to this work.}}
\author{Xufei Suo\textsuperscript{1}}
\author{Feng Jiang}
\author{Yongxing Shen\corref{correspondingauthor}}
\cortext[correspondingauthor]{Corresponding author}
\address{University of Michigan -- Shanghai Jiao Tong University Joint Institute, Shanghai Jiao Tong University, Shanghai, China, 200240}
\ead{yongxing.shen@sjtu.edu.cn}

\begin{abstract}
The mechanical properties of metal matrix fiber-reinforced composites depend on many aspects of their structure in a complicated way. In this paper, we propose a \emph{minimalistic} approach to study interface debonding, matrix cracking, and their competition in metal matrix fiber-reinforced elastoplastic composites by numerical simulation. This approach combines a cohesive zone model for interface debonding and a phase field model for matrix cracking. The features of this framework are: (1) crack nucleation, propagation, and branching can be easily tracked without the need of geometric programming; (2) the interface debonding is determined merely by the CZM, but not interfered by the phase field in the bulk; (3) the cohesive interface has zero thickness instead of being regularized; (4) any reasonable cohesive law of interest is readily incorporated with very few constraints; (5) elastoplasticity of the matrix is conveniently taken into account, as strains in the model are all well defined; (6) the competition of the two failure mechanisms, namely, matrix cracking and interface debonding, is accurately captured. Accuracy of this framework is verified with existing analytical and numerical results. The proposed framework shows a potential in investigating various complicated crack behaviors in composites.
\end{abstract}

\begin{keyword}
Metal matrix fiber-reinforced composites\sep Cohesive zone model\sep Phase field model\sep Interface debonding\sep Matrix cracking
\end{keyword}

\end{frontmatter}

\section{Introduction}\label{intro}
With the rapid development of the aerospace industry, there is an increasing demand for improving the performance of structural materials. In recent decades, metal matrix fiber-reinforced composites have attracted much attention and achieved wide applications due to their excellent properties, including high strength, high stiffness, and temperature resistance. These advantages guarantee the high-quality in manufacturing of aerospace components such as aircraft skins and integral blade rings.

It is well known that preventing material failure in composites is a significant issue. However, due to high manufacturing cost, it is expensive and time-consuming to determine the reliability of composites through experiments. Therefore, numerical simulations of microscopic characteristics of composites become necessary and important. In recent decades, various computational methods for modeling fracture behaviors have been developed. One widely used method is the cohesive zone model (CZM) \cite{park2011cohesive,dugdale1960yielding,barenblatt1962mathematical}, which is based on the cohesive interactions, or so-called traction-separation law, to simulate progressive nonlinear fracture behaviors. In the CZM, the crack path highly relies on the location of specified cohesive elements. Some applications of the CZM for failure analysis in composites are in \cite{CZM1997,CZM2015,CZM-XFEM}. Another very successful method, the extended finite element method (XFEM) \cite{moes1999finite}, is able to simulate crack propagation without remeshing by adding discontinuous enrichment functions to account for the existing cracks. Some applications of the XFEM for failure analysis in composites are performed in \cite{CZM-XFEM,XFEM2003,XFEM2004,XFEM2009,XFEM2013,XFEM2017}.

Recently, the phase field model (PFM), a variational approach to fracture, has become prevalent in computational fracture mechanics. Originated from the variational formulation of brittle fracture by Francfort and Marigo \cite{Francfort1998-26}, the PFM was proposed by Bourdin et al.~\cite{Bourdin2000-24} and has advantages in simulating crack initiation, propagation and branching without tracking the crack path, features not possessed by the CZM and the XFEM. The PFM introduces a scalar variable between 0 and 1 indicating unbroken and broken regions, respectively, to describe the damage of the material. An overview of the PFM and detailed implementation of it are presented in \cite{PFM-review,Shen2018-27,PFM-review-Xiaofei}.

\begin{figure}[htbp]
	\centering
	\includegraphics[width=0.4\textwidth]{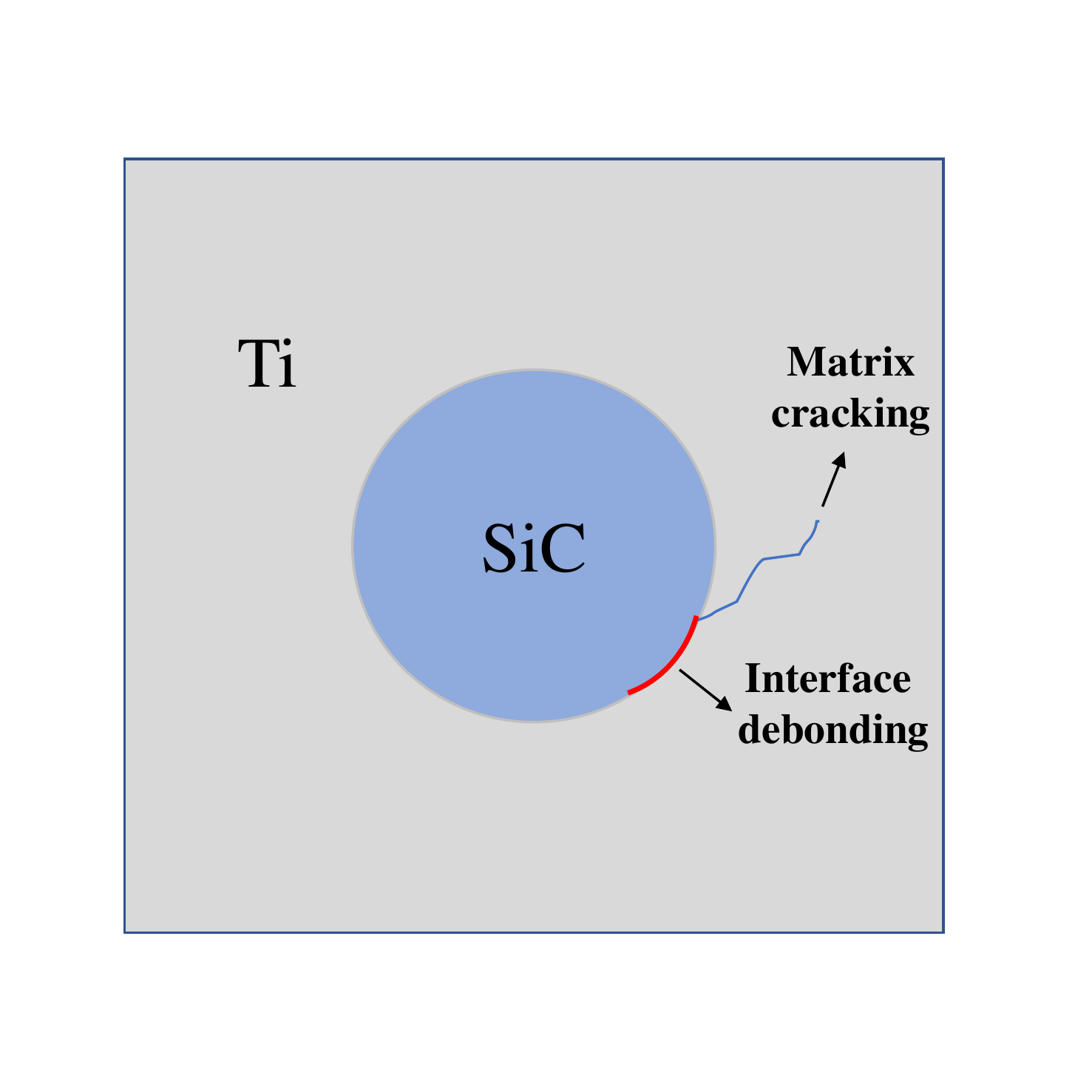}
	\caption{Representative failure mechanisms of metal matrix fiber-reinforced composites.}
	\label{failure-mechanism}
\end{figure}

Over the past few years, the PFM has been used to investigate the progressive failure of fiber-reinforced composites by simultaneously considering two major failure mechanisms, matrix cracking and interface debonding, as illustrated in Fig.~\ref{failure-mechanism}. Most of these references investigated brittle fracture in fiber-reinforced composites \cite{nguyen2016phase,Reddy2021,ZhangP2019,LW2020,Tarafder2020,paggi2017revisiting,zhang2020model,R-curve2021}  while Ref.~\cite{Pengfei2020} extended this investigation to elastoplastic composites. These contributions can be categorized into two families based on how the interface is treated and how debonding is simulated. 

The first family regularizes the interface over a finite width and represents the interface failure with an auxiliary interface phase field \cite{nguyen2016phase,Reddy2021,ZhangP2019,LW2020,Tarafder2020,Pengfei2020}. Such a treatment leads to a diffusive interface rather than a sharp one, thus demanding extra studies of the effect of the interface regularization length scale. One major difference among these works concerns the form of the interface fracture energy density: in \cite{nguyen2016phase,Reddy2021,Tarafder2020,Pengfei2020}, the interface fracture energy density is directly adopted from the CZM; while in \cite{ZhangP2019,LW2020}, this quantity is replaced by a modified interface fracture toughness to account for the influences of the fiber and the matrix on interface failure.

The second family models the interface to have zero thickness and adopts the CZM for interface debonding \cite{paggi2017revisiting,zhang2020model,R-curve2021}. Such zero-thickness cohesive elements eliminate the need of regularizing the interface, of solving for the interface phase field, and of studying the effect of the associated regularization length scale. In particular, Paggi and Reinoso \cite{paggi2017revisiting} assumed the critical opening displacement of the adopted linear CZM is dependent on the phase field in the surrounding matrix; Zhang et al.~\cite{zhang2020model} introduced an equivalent damage index as the biggest damage state among that of the cohesive elements and of their adjacent bulk elements. As a result, in both models, the interface failure depends not only on the CZM but also on the phase field of the adjacent bulk element.

In this work, we propose a \emph{minimalistic} hybrid framework, to simulate two failure mechanisms in metal matrix fiber-reinforced composites, in the hope of elucidating how each feature contributes to the overall behavior of the composite. In particular, we utilize the PFM for matrix cracking and the CZM for interface debonding, and adopt zero-thickness cohesive elements as a sharp interface. 
As a result, the proposed framework also enjoys the advantages of the second family mentioned above.
Moreover, the interface debonding is simulated merely based on the CZM and is not interfered by the phase field in the bulk.

For the problem at hand, the proposed framework is also advantageous in a few aspects. First of all, the elastoplasticity of the matrix is conveniently taken into account. This is not so straightforward for methods in the first family. For example, in \cite{Pengfei2020}, the total strain is the sum of the elastic strain, the plastic strain, and an equivalent strain corresponding to the smear interface. As the last term has an artificial parameter, the tracking of elastoplasticity therein is ambiguous.

In addition, any reasonable cohesive law can be used with hardly any restriction, unlike the model of \cite{paggi2017revisiting}.

Even with such minimum setting, the proposed framework is capable of accurately capturing the competition of the two failure mechanisms, i.e., matrix cracking and interface debonding.

The structure of this paper is arranged as follows. Section \ref{main} details the variational formulation of the proposed framework and the numerical aspect, along with a comparison with existing works. Section \ref{example} shows the performance of the model through different numerical applications. Finally, Section \ref{conclusion} summarizes the main features of the proposed framework.

\section{Phase field formulation for composite fracture with two failure mechanisms}\label{main}

In this section, the mathematical model for two failure mechanisms, i.e., matrix cracking and interface debonding, of metal matrix fiber-reinforced elasotplastic composites is detailed, which has the feature of having the \emph{minimum} ingredient compared with existing models for similar purposes in
\cite{nguyen2016phase,Pengfei2020,Tarafder2020,Reddy2021,ZhangP2019,LW2020,paggi2017revisiting,zhang2020model,R-curve2021}. The overall potential energy functional is written as
\begin{equation}\label{total}
    \Pi=\Pi_\mathrm{e}+\Pi_\mathrm{p}+\Pi_\mathrm{f}+\Pi_\mathrm{i}-\Pi_\mathrm{w}
\end{equation}
where $\Pi_\mathrm{e}$, $\Pi_\mathrm{p}$, $\Pi_\mathrm{f}$, $\Pi_\mathrm{i}$, and $\Pi_\mathrm{w}$ are the elastic strain energy, the plastic strain energy, the fracture surface energy, the interface energy due to relative displacement of the two sides, and the external work, respectively. Each term on the right hand side of Eq.~\eqref{total} is explained in the sequel. To be specific, in Section \ref{phase field}, the phase field model (PFM) for ductile matrix cracking with specific plastic evolution law is introduced. Then, Section \ref{debonding} incorporates interface debonding by combining the PFM with the cohesive zone model (CZM). Section \ref{discretization} is devoted to the finite element discretization of the proposed framework, followed by a comparison with existing models in Section \ref{sec:rel-ex-mod}.

\subsection{Ductile matrix cracking}\label{phase field}

The PFM is utilized to model ductile matrix cracking. For the ductile fracture of a single material, Eq.~\eqref{total} is specialized to
\begin{equation}\label{ductile-phase field}
    \Pi[\boldsymbol{u},d;\boldsymbol{\varepsilon}_{\mathrm{p}};e_{\mathrm{p}}]=\Pi_\mathrm{e}[\boldsymbol{u},d;\boldsymbol{\varepsilon}_{\mathrm{p}}]+\Pi_\mathrm{p}[e_{\mathrm{p}}]+\Pi_\mathrm{f}[d]-\Pi_\mathrm{w}[\boldsymbol{u}],
\end{equation}
where $\boldsymbol{u}$ and $d$ denote the displacement field and the phase field. The phase field $d$, ranging from 0 to 1, is a diffuse field introduced to represent the cracked material. Wherever $d=0$, the material is intact and wherever $d=1$, the material is fully broken. Also, $\boldsymbol{\varepsilon}_{\mathrm{p}}=\boldsymbol{\varepsilon}-\boldsymbol{\varepsilon}_{\mathrm{e}}$ is the plastic strain, in which $\boldsymbol{\varepsilon}$, the total strain, is defined as $\boldsymbol{\varepsilon}:= \left[\nabla \boldsymbol{u} + \left(\nabla \boldsymbol{u}\right)^{\rm{T}} \right]/2$ and $\boldsymbol{\varepsilon}_{\mathrm{e}}$ is the elastic strain. $e_{\mathrm{p}}$ represents the accumulation effect of $\boldsymbol{\varepsilon}_{\mathrm{p}}$.

Let $\Omega \subset \mathbb{R}^{n}$, $n=2$ or 3, represent the space occupied by the solid of interest in its undeformed state. Its Lipschitz-continuous boundary, $\partial \Omega$, is made of two mutually disjoint subsets $\Gamma_D$ and $\Gamma_N$, satisfying $\Gamma_{D} \cup \Gamma_{N}=\partial \Omega$ and $\Gamma_{D} \cap \Gamma_{N}=\emptyset$. For $\boldsymbol{u} \in H^1(\Omega,\mathbb{R}^{n})$ and $d \in H^1(\Omega)\cap L^\infty(\Omega)$, the terms on the right hand side of Eq.~\eqref{ductile-phase field} are given by
\begin{equation}\label{pi-form}
\begin{aligned}
\Pi_\mathrm{e}[\boldsymbol{u},d;\boldsymbol{\varepsilon}_{\mathrm{p}}]&=\int_{\Omega} \Psi_\mathrm{e}[\boldsymbol{\varepsilon}_{\mathrm{e}}(\boldsymbol{u}), d] \;\mathrm{d}\Omega,\\
\Pi_\mathrm{p}[e_{\mathrm{p}}]&=\int_{\Omega} \Psi_\mathrm{p}[e_{\mathrm{p}}] \;\mathrm{d}\Omega,\\
\Pi_\mathrm{f}[d]&=g_{c} \int_{\Omega} \gamma(d, \nabla d)\; \mathrm{d} \Omega,\\
\Pi_\mathrm{w}[\boldsymbol{u}]&=\int_{\Gamma_{N}} \boldsymbol{t}_{N} \cdot \boldsymbol{u}\; \mathrm{d} \Gamma+\int_{\Omega} \boldsymbol{b} \cdot \boldsymbol{u}\; \mathrm{d} \Omega,
\end{aligned}  
\end{equation}
where
\begin{equation}\label{gamma-bulk}
  \gamma(d, \nabla d)=\frac{d^{2}}{2 \ell}+\frac{\ell}{2}|\nabla d|^{2}  
\end{equation}
is the crack surface density per unit volume, in which scalar parameters $\ell$ and $g_c$ are the regularization length scale and critical energy release rate of crack propagation, respectively. Scalar functions $\Psi_\mathrm{e}$ and $\Psi_\mathrm{p}$ are the elastic and plastic strain energy densities, respectively. Vector fields $\boldsymbol{t}_{N}: \Gamma_{N} \rightarrow \mathbb{R}^{n}$, $\boldsymbol{b}: \Omega \rightarrow \mathbb{R}^{n}$, and $\boldsymbol{u}_{D}: \Gamma_{D} \rightarrow \mathbb{R}^{n}$ are the prescribed traction, body force, and boundary displacement, respectively. To account for the tension-compression asymmetry, $\Psi_\mathrm{e}(\boldsymbol{\varepsilon}_{\mathrm{e}},d)$ takes the following form
$$\Psi_\mathrm{e}(\boldsymbol{\varepsilon}_{\mathrm{e}},d)=g(d)\Psi_{+}(\boldsymbol{\varepsilon}_{\mathrm{e}})+\Psi_{-}(\boldsymbol{\varepsilon}_{\mathrm{e}}),
$$
where $g(d)=(1-d)^{2}+k$ is the degradation function with $k$ a small positive number to avoid a singular tangent stiffness matrix when the material is fully broken, and $\Psi_{+}(\boldsymbol{\varepsilon}_{\mathrm{e}})$ and $\Psi_{-}(\boldsymbol{\varepsilon}_{\mathrm{e}})$ account for the crack-driving and persistent parts of the strain energy density, respectively. In this work the volumetric-deviatoric decomposition proposed by Amor et al.~\cite{amor2009regularized} is adopted among the different options in the literature, such that
\begin{equation}
\label{update-Psi}
   \Psi_{+}(\boldsymbol{\varepsilon}_{\mathrm{e}})=\frac{\kappa}{2}\langle\operatorname{tr} \boldsymbol{\varepsilon}_{\mathrm{e}}\rangle_{+}^{2}+\mu\|\operatorname{dev} \boldsymbol{\varepsilon}_{\mathrm{e}}\|^{2}, \quad
		\Psi_{-}(\boldsymbol{\varepsilon}_{\mathrm{e}})=\frac{\kappa}{2}\langle\operatorname{tr} \boldsymbol{\varepsilon}_{\mathrm{e}}\rangle_{-}^{2}, 
\end{equation}
where $\kappa>0$ and $\mu>0$ are the bulk modulus and shear modulus, respectively, $\operatorname{dev}\boldsymbol{\varepsilon}_{\mathrm{e}}=\boldsymbol{\varepsilon}_{\mathrm{e}}-(1/3)(\operatorname{tr}\boldsymbol{\varepsilon}_{\mathrm{e}})\boldsymbol{1}$, $\boldsymbol{1}$ the second-order identity tensor, $\langle a\rangle_{\pm}:=(a \pm|a|)/2$, $\|\cdot\|:\boldsymbol{A}\mapsto\sqrt{\boldsymbol{A}:\boldsymbol{A}}$ is the Frobenius norm, and the Cauchy stress tensor is given by
\begin{equation}\label{ductile-stress}
	\begin{aligned}
		\boldsymbol{\sigma}(\boldsymbol{\varepsilon}_{\mathrm{e}}, d)=\frac{\partial \Psi_\mathrm{e}(\boldsymbol{\varepsilon}_{\mathrm{e}}, d)}{\partial \boldsymbol{\varepsilon}_{\mathrm{e}}} =g(d)\left(\kappa\langle\operatorname{tr} \boldsymbol{\varepsilon}_{\mathrm{e}}\rangle_{+}\boldsymbol{1}+2 \mu \operatorname{dev} \boldsymbol{\varepsilon}_{\mathrm{e}}\right)+\kappa\langle\operatorname{tr} \boldsymbol{\varepsilon}_{\mathrm{e}}\rangle_{-}\boldsymbol{1}.
	\end{aligned}
\end{equation}

\paragraph{Equilibrium equations for the fields}
Equilibrium is obtained by taking the first variation of Eq.~\eqref{ductile-phase field} with respect to $\boldsymbol{u}$, which gives
\begin{equation}\label{ductile-weakform-u}
\int_{\Omega} \boldsymbol{\sigma}[\boldsymbol{\varepsilon}_{\mathrm{e}}(\boldsymbol{u}), d] : \boldsymbol{\varepsilon}_{\mathrm{e}}(\overline{\boldsymbol{u}}) \; \mathrm{d} \Omega-\int_{\Gamma_{N}} \boldsymbol{t}_{N} \cdot \overline{\boldsymbol{u}} \; \mathrm{d} \Gamma-\int_{\Omega} \boldsymbol{b} \cdot \overline{\boldsymbol{u}} \; \mathrm{d} \Omega=0,\quad
\forall \overline{\boldsymbol{u}}\in H^1(\Omega,\mathbb{R}^{n}),\overline{\boldsymbol{u}}=\boldsymbol{0}\text{ on }\Gamma_D,
\end{equation}
and by taking the first variation of Eq.~\eqref{ductile-phase field} with respect to $d$, which yields
\begin{equation}\label{ductile-weakform-d}
\int_{\Omega} g^{\prime}(d) \Psi_{+}(\boldsymbol{\varepsilon}_{\mathrm{e}}) \overline{d} \; \mathrm{d} \Omega+g_c \int_{\Omega}\left[\frac{d \overline{d}}{\ell}+\ell \nabla d \cdot \nabla\overline{d}\right] \mathrm{d} \Omega=0,\quad\forall \overline{d} \in H^1(\Omega)\cap L^\infty(\Omega).
\end{equation}

In addition, to enforce the irreversibility of the phase field, a so-called history parameter $\mathcal{H}$ is introduced following the idea of \cite{miehe2010phase}:
$$
    \mathcal{H}(\boldsymbol{x},t)=\max_{0 \leq \tau \leq t} \Psi_{+}(\boldsymbol{x},\tau),\quad \boldsymbol{x} \in \Omega.
$$
Eq.~\eqref{ductile-weakform-d} is then replaced by
\begin{equation}\label{ductile-weakform-d-H}
    \int_{\Omega} g^{\prime}(d) \mathcal{H} \overline{d} \; \mathrm{d} \Omega+g_c \int_{\Omega}\left[\frac{d \overline{d}}{\ell}+\ell \nabla d \cdot \nabla\overline{d}\right] \mathrm{d} \Omega=0,\quad\forall \overline{d} \in H^1(\Omega)\cap L^\infty(\Omega).
\end{equation}

\paragraph{Plastic constitutive model}
In this work, we adopt the isotropic hardening model with the yield function expressed as \cite{simocomputational}
$$
f\left(\boldsymbol{\mathrm{s}}, e_{\mathrm{p}}\right)=
\|\boldsymbol{\mathrm{s}}\|-\sqrt{\frac{2}{3}}\left[\sigma_{\mathrm{Y}}^{0}+ K e_{\mathrm{p}}\right],
$$
where $\boldsymbol{\mathrm{s}}=\operatorname{dev}\boldsymbol{\sigma}$ is the deviatoric stress, $e_{\mathrm{p}}$ is the effective plastic strain with $\dot{e}_{\mathrm{p}}=\sqrt{\left(2/3\right) \dot{\boldsymbol{\varepsilon}}_{\mathrm{p}}: \dot{\boldsymbol{\varepsilon}}_{\mathrm{p}}}$, $\sigma_{\mathrm{Y}}^{0}$ is the initial yield stress, and $K$ is the plastic modulus. The Kuhn–Tucker condition and plastic consistency condition \cite{simocomputational} read
$$
	\begin{array}{c}
		f\left(\boldsymbol{\mathrm{s}}, e_{\mathrm{p}}\right) \leq 0, \quad \dot{e}_\mathrm{p} \geq 0, \quad \dot{e}_\mathrm{p} f\left(\boldsymbol{\mathrm{s}}, e_{\mathrm{p}}\right)=0, \quad \dot{e}_\mathrm{p} \dot{f}\left(\boldsymbol{\mathrm{s}}, e_{\mathrm{p}}\right)=0.
	\end{array}
$$

\subsection{Interface debonding}\label{debonding}
In this subsection, interface debonding is incorporated with irreversibility to account for the weakening effect of tension loading. For this purpose, Eq.~\eqref{total} is rewritten as
\begin{equation}\label{cohesive-phase field}
    \Pi[\boldsymbol{u},d;\boldsymbol{\varepsilon}_{\mathrm{p}};e_{\mathrm{p}};\Delta_{\max}]=\Pi_\mathrm{e}[\boldsymbol{u},d;\boldsymbol{\varepsilon}_{\mathrm{p}}]+\Pi_\mathrm{p}[e_{\mathrm{p}}]+\Pi_\mathrm{f}[d]+\Pi_\mathrm{i}[\boldsymbol{u};\Delta_{\max}]-\Pi_\mathrm{w}[\boldsymbol{u}],
\end{equation}
where $\Pi_\mathrm{e}$, $\Pi_\mathrm{f}$, and $\Pi_\mathrm{w}$ take the same form as in Eq.~\eqref{pi-form}, while $\Delta_{\max}$ is the history parameter for irreversibility to be defined in the sequel. For $\boldsymbol{u} \in H^1(\Omega,\mathbb{R}^{n})$, $d \in H^1(\Omega)\cap L^\infty(\Omega)$, and $\Delta_{\max} \in H^{1/2}(\Gamma_I)$, $\Pi_\mathrm{i}$ is given by
$$
\Pi_\mathrm{i}[\boldsymbol{u};\Delta_{\max}]=\int_{\Gamma_{I}} \phi \; \mathrm{d} \Gamma,
$$
where $\Gamma_{I}$ is the interface and $\phi$ represents the interface fracture energy per unit area. Here in the initial loading path, the form of $\phi$ is adopted from Xu and Needleman \cite[Equations (4.7) and (4.8)]{xu1993void} with simplification of $q=1$ \cite{improve-xu}, giving
\begin{equation}\label{exponential}
    \phi^\mathrm{initial}\left(\mathbf{\Delta}\right)=\phi_{\mathrm{n}}+\phi_{\mathrm{n}} \exp \left(-\frac{\Delta_{\mathrm{n}}}{\delta_{\mathrm{n}}}\right)\left[-\left(1+\frac{\Delta_{\mathrm{n}}}{\delta_{\mathrm{n}}} \right) \exp \left(-\frac{\Delta_{\mathrm{t}}^{2}}{\delta_{\mathrm{t}}^{2}}\right)\right],
\end{equation}
where $\mathbf{\Delta}=\left\{\Delta_{\mathrm{n}}, \Delta_{\mathrm{t}}\right\}$ with $\Delta_{\mathrm{n}}$ and $\Delta_{\mathrm{t}}$ the current normal and tangential separations, respectively; $\delta_{\mathrm{n}}$ and $\delta_{\mathrm{t}}$ are material parameters called characteristic normal and tangential separations, respectively. Here the relation between the interface fracture energy $\phi_{\mathrm{n}}$ and $\phi_{\mathrm{t}}$ and the characteristic separations $\delta_{\mathrm{n}}$ and $\delta_{\mathrm{t}}$ are related by
$$
\phi_{\mathrm{n}}=e \sigma_{\mathrm{max}}\delta_{\mathrm{n}},
\quad
\phi_{\mathrm{t}}=\sqrt{\frac{e}{2}} \tau_{\mathrm{max}}\delta_{\mathrm{t}},
$$
where $e=\exp(1)$ is the base of the natural logarithm; $\sigma_{\mathrm{max}}$ and $\tau_{\mathrm{max}}$ are the maximum normal and shear strengths, respectively.

\begin{figure}[htbp]
	\centering
	\includegraphics[width=0.7\textwidth]{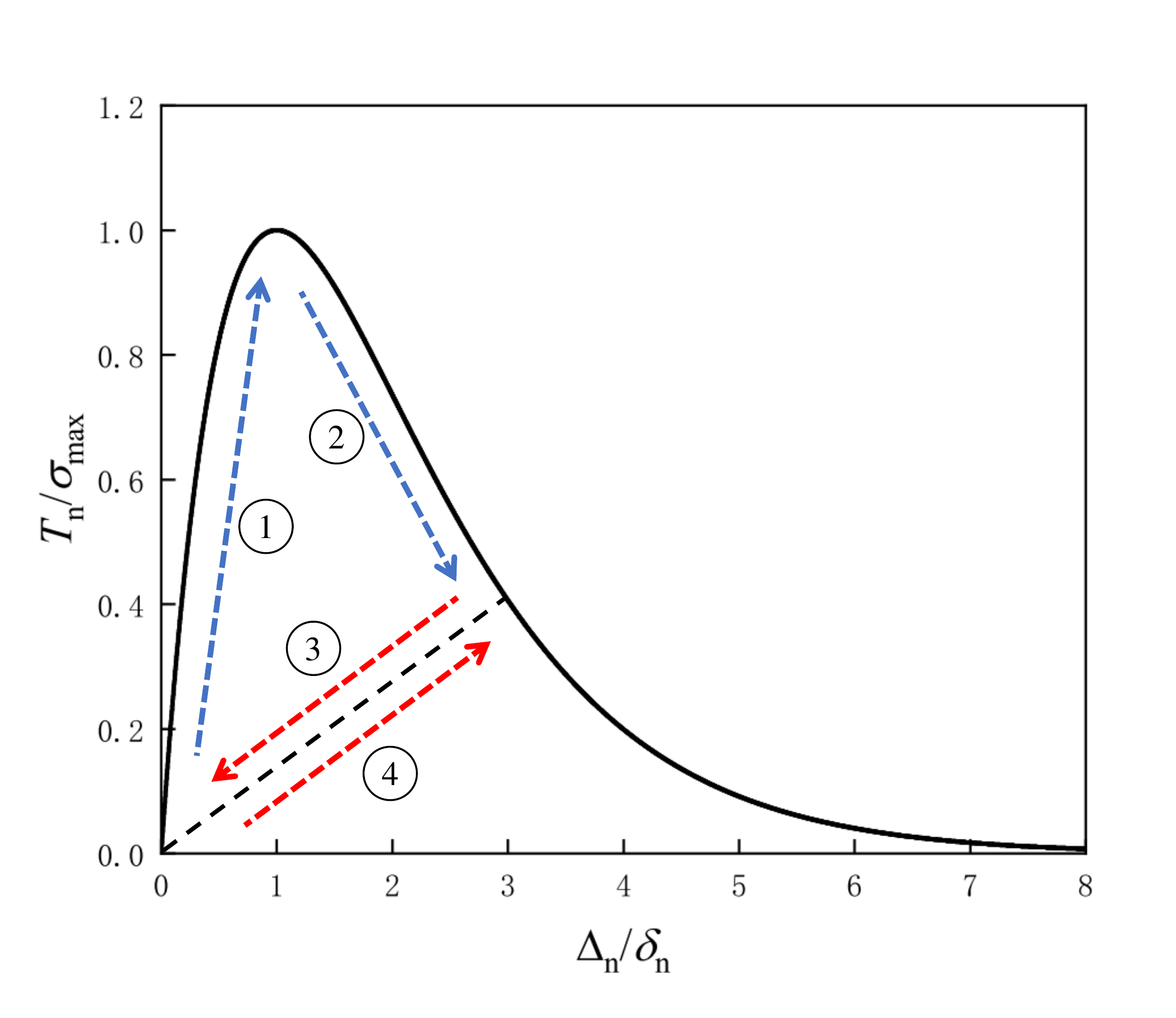}
	\caption{Exponential cohesive law adopted from Xu and Needleman \cite{xu1993void} for loading (black solid curve) and its extension to unloading \ding{174} or reloading \ding{175} (red dash line).}
	\label{unload}
\end{figure}

Irreversibility of interface debonding is incorporated in the CZM as shown in Figure \ref{unload}. Once any point on the interface is loaded in tension for the first time, its traction follows the black solid curve and the interface is irreversibly weakened; whereas the same point is unloaded or reloaded, its traction follows the red dash line. More precisely, the traction at any point of the cohesive interface is given by
\begin{equation}\label{TNTT}
\boldsymbol{T}=
\begin{cases}
-\dfrac{{\partial \phi^\mathrm{initial}(\mathbf{\Delta})}}{{\partial \mathbf{\Delta}}} & \text{if $\Delta=\Delta_{\max}$ and $\dot{\Delta} \geq 0$},\\
-\dfrac{1}{\hat{\Delta}}\dfrac{{\partial \phi^\mathrm{initial}(\hat{\Delta}\mathbf{\Delta})}}{{\partial\mathbf{\Delta}}} & \text{otherwise},
\end{cases}
\end{equation}
where $\boldsymbol{T}=\left\{T_{\mathrm{n}}, T_{\mathrm{t}}\right\}$ with $T_{\mathrm{n}}$ and $T_{\mathrm{t}}$ the current normal and tangential traction, respectively, and $\Delta_{\max}$ is introduced to account for the irreversibility of interface debonding \cite{ortiz1999finite},
$$
    \Delta_{\max}(\boldsymbol{x},t)=\max_{0 \leq \tau \leq t} \Delta(\boldsymbol{x},\tau),\quad \boldsymbol{x} \in \Gamma_{I},
$$
with $\Delta=\sqrt{\Delta_{\mathrm{n}}^{2}+\Delta_{\mathrm{t}}^{2}}$ and $\hat{\Delta}=\Delta_{\max} / \Delta$.

Here a simplificaton is made to ease the subsequent calculation. Whenever the normal separation at any point is such that $\Delta_{\mathrm{n}}>6\delta_{\mathrm{n}}$, its normal traction is set to zero ever since.

Herein, the weak form for the displacement is written as
\begin{equation}\label{all-delta-u}
    \int_{\Omega} \boldsymbol{\sigma}[\boldsymbol{\varepsilon}_{\mathrm{e}}(\boldsymbol{u}), d] : \boldsymbol{\varepsilon}_{\mathrm{e}}(\overline{\boldsymbol{u}}) \; \mathrm{d} \Omega-\int_{\Gamma_{I}} \boldsymbol{T} \cdot \overline{\mathbf{\Delta}} \; \mathrm{d} S-\int_{\Gamma_{N}} \boldsymbol{t}_{N} \cdot \overline{\boldsymbol{u}} \; \mathrm{d} \Gamma-\int_{\Omega} \boldsymbol{b} \cdot \overline{\boldsymbol{u}} \; \mathrm{d} \Omega=0,
\end{equation}
where $\overline{\mathbf{\Delta}}$ is the virtual separation, which is a linear functional of the value of $\overline{\boldsymbol{u}}$ restricted to both sides of $\Gamma_{I}$. Compared with Eq.~\eqref{ductile-weakform-u}, because of the additional term $\Pi_\mathrm{i}$ in Eq.~\eqref{cohesive-phase field}, Eq.~\eqref{all-delta-u} adds the first variation of $\Pi_\mathrm{i}$ with respect to $\boldsymbol{u}$.

\subsection{Finite element discretization}\label{discretization}

For convenience, in this subsection, we will confine ourselves to the plane strain case, while the formulation is applicable to three dimensional cases with minor changes.

The finite element discretization is presented in this subsection. The displacement field $\boldsymbol{u}$ and the phase field $d$ are discretized as:
\begin{equation}
		\boldsymbol{u}(\boldsymbol{x})=\sum\limits_{i=1}^{n}\boldsymbol{N}_i(\boldsymbol{x})\mathbf{u}_i, \quad d(\boldsymbol{x})=\sum\limits_{i=1}^{n}N_i(\boldsymbol{x})d_i,
\end{equation}
where $\mathbf{u}_i$ and $d_i$ are the displacement vector and phase field values at node $i$, respectively; $N_i$ is the standard Lagrangian shape functions associated with node $i$ and
$$
\boldsymbol{N}_i=
\begin{bmatrix}
N_i & 0 \\
0 & N_i
\end{bmatrix}.
$$

For convenience in the sequel, we define $\mathbf{u}$ and $\mathbf{d}$ as column vectors containing all entries of $\mathbf{u}_i$ and $d_i$, respectively. According to Eq.~\eqref{all-delta-u} and Eq.~\eqref{ductile-weakform-d-H}, the residual vectors associated with node $i$ read
\begin{align}\label{discretization-Ru}
\mathbf{R}_i^{\mathbf{u}}&=\int_{\Omega} \boldsymbol{B}_i^{\rm{T}} \boldsymbol{\sigma}[\boldsymbol{\varepsilon}_{\mathrm{e}}(\boldsymbol{u}), d]\; \mathrm{d} \Omega-\int_{\Gamma_{I}} \boldsymbol{L}_i^{\rm{T}} \boldsymbol{T} \; \mathrm{d} S-\int_{\Gamma_{N}} \boldsymbol{N}_i \boldsymbol{t}_{N} \; \mathrm{d} \Gamma-\int_{\Omega} \boldsymbol{N}_i \boldsymbol{b} \; \mathrm{d} \Omega,\\
\label{discretization-Rd}
\mathbf{R}_i^{\mathbf{d}}&=\int_{\Omega} \left\{\left[\frac{g_c}{\ell} d+g^{\prime}(d)\mathcal{H}\right]N_i+g_c \ell \nabla d \cdot \nabla N_i\right\}\; \mathrm{d} \Omega,
\end{align}
where $\boldsymbol{L}_i$ is the displacement-separation matrix for node $i$ such that $\boldsymbol{\Delta}_i=\boldsymbol{L}_i\mathbf{u}$ \cite{cohesive-transform}, $\boldsymbol{B}_i$ is the strain-displacement matrix for node $i$:
$$
\boldsymbol{B}_i=
\begin{bmatrix}
N_{i,x} & 0 \\
0 & N_{i,y} \\
N_{i,y} & N_{i,x}
\end{bmatrix}.
$$

In a staggered solution scheme, the entries of the corresponding tangent stiffness matrices read
\begin{align}\label{discretization-Ku}
\mathbf{K}_{ij}^{\mathbf{u}}&=\int_{\Omega} \boldsymbol{B}_i^{\rm{T}}\boldsymbol{D}\boldsymbol{B}_j\; \mathrm{d} \Omega-\int_{\Gamma_{I}} \boldsymbol{L}_i\frac{\partial\boldsymbol{T}}{\partial\mathbf{\Delta}} \boldsymbol{L}_j\; \mathrm{d} S,\\\label{discretization-Kd}
\mathbf{K}_{ij}^{\mathbf{d}}&=\int_{\Omega} \left[\left(g^{\prime\prime}(d)\mathcal{H}+\frac{g_c}{\ell}\right)N_i N_j+g_c \ell\nabla N_i \cdot \nabla N_j\right] \mathrm{d} \Omega,
\end{align}
where
 $\partial\boldsymbol{T} / \partial\mathbf{\Delta}$ is the $n\times n$ stiffness matrix for the interface cohesive elements and $\boldsymbol{D}$ is the stiffness matrix for the bulk elements, given by $\boldsymbol{D}=g(d)\boldsymbol{D}_+ +\boldsymbol{D}_-$, where $\boldsymbol{D}_+$ and $\boldsymbol{D}_-$ are given by \cite{Shen2018-27}
$$
\begin{aligned}
\boldsymbol{D}_+=\kappa \textup{H}(\operatorname{tr}\boldsymbol{\varepsilon}_{\mathrm{e}})
\begin{bmatrix}
1 & 1 & 0 \\
1 & 1 & 0 \\
0 & 0 & 0
\end{bmatrix}
+\frac{\mu}{3}
\begin{bmatrix}
4 & -2 & 0 \\
-2 & 4 & 0 \\
0 & 0 & 3
\end{bmatrix},
\quad
\boldsymbol{D}_-=\kappa \textup{H}(-\operatorname{tr}\boldsymbol{\varepsilon}_{\mathrm{e}})
\begin{bmatrix}
1 & 1 & 0 \\
1 & 1 & 0 \\
0 & 0 & 0
\end{bmatrix},
\end{aligned}
$$
where $\textup{H}$ is the Heaviside function such that $\textup{H}(a)=1$ if $a>0$, $\textup{H}(a)=0$ if $a<0$, and $\textup{H}(a)=1/2$ if $a=0$ for symmetry. The overall algorithm is described in Algorithm \ref{algorithm-total}.

\begin{algorithm}[htbp]
\caption{Algorithm for modeling ductile matrix cracking and interface debonding.}\label{algorithm-total}
\SetKwInOut{Input}{Input}
\SetKwInOut{Output}{Output}
\Input{$\mathbf{u}_{m}$, the nodal displacement vector at the $m$th step; $\mathbf{d}_{m}$, the nodal phase field vector at the $m$th step; $\boldsymbol{\varepsilon}_{\mathrm{p}}^m$, the plastic strain at the $m$th step; $\mathcal{H}_m$ and $\Delta_{\max}^m$, history variables at the $m$th step; $\epsilon_{\mathrm{tol}}^{\mathbf{u}}$, the tolerance for $\mathbf{R}_{\mathbf{u}}$; $\epsilon_{\mathrm{tol}}^{\mathbf{d}}$ the tolerance for $\mathbf{R}_\mathbf{d}$; material properties and boundary conditions needed to compute the residuals and tangent stiffnesses}
\Output{$\mathbf{u}_{m+1}$; $\mathbf{d}_{m+1}$; $\boldsymbol{\varepsilon}_{\mathrm{p}}^{m+1}$; $\mathcal{H}_{m+1}$; $\Delta_{\max}^{m+1}$}
$\mathbf{u}\leftarrow\mathbf{u}_m$, $\mathbf{d}\leftarrow\mathbf{d}_m$, $\mathcal{H}\leftarrow\mathcal{H}_m$, $\Delta_{\max}\leftarrow\Delta_{\max}^m$\;\tcc{Making a copy of the last converged results}
Compute $\mathbf{R}_{\mathbf{u}}=\mathbf{R}_{\mathbf{u}}(\mathbf{u},\mathbf{d},\Delta_{\max})$ with Eq.~\eqref{all-delta-u}\;
\tcc{Whenever the function $\mathbf{R}_{\mathbf{u}}(\cdot)$ is invoked, $\boldsymbol{\varepsilon}_\mathrm{p}$ is updated}
Compute $\mathbf{R}_\mathbf{d}=\mathbf{R}_\mathbf{d}(\mathbf{u},\mathbf{d},\mathcal{H})$ with Eq.~\eqref{ductile-weakform-d-H}\;
\While{$\mathbf{R}_{\mathbf{u}}>\epsilon_{\mathrm{tol}}^{\mathbf{u}}$ or $\mathbf{R}_\mathbf{d}>\epsilon_{\mathrm{tol}}^{\mathbf{d}}$}{
\While{$\mathbf{R}_{\mathbf{u}}>\epsilon_{\mathrm{tol}}^{\mathbf{u}}$}{
Compute $\mathbf{K}_{\mathbf{u}}=\mathbf{K}_{\mathbf{u}}(\mathbf{u},\mathbf{d},\Delta_{\max})$ with Eq.~\eqref{all-delta-u}\;
$\mathbf{u}\leftarrow \mathbf{u}-\mathbf{K}_{\mathbf{u}}^{-1}\mathbf{R}_{\mathbf{u}}$\;
Compute $\Psi_{+}(\mathbf{u})$ with Eq.~\eqref{update-Psi}\;
$\mathcal{H} = \max\{\Psi_{+},\mathcal{H}_m\}$\;
Compute $\Delta$ from $\mathbf{u}$\;
$\Delta_{\max} = \max\{\Delta,\Delta_{\max}^m\}$\;
Update $\mathbf{R}_{\mathbf{u}}=\mathbf{R}_{\mathbf{u}}(\mathbf{u},\mathbf{d},\Delta_{\max})$\;
}
Update $\mathbf{R}_{\mathbf{d}}=\mathbf{R}_{\mathbf{d}}(\mathbf{u},\mathbf{d},\mathcal{H})$\;
\While{$\mathbf{R}_{\mathbf{d}}>\epsilon_{\mathrm{tol}}^{\mathbf{d}}$}{
Compute $\mathbf{K}_\mathbf{d}(\mathbf{u},\mathbf{d},\mathcal{H})$ with Eq.~\eqref{ductile-weakform-d-H}\;
$\mathbf{d}\leftarrow \mathbf{d}-\mathbf{K}_{\mathbf{d}}^{-1}\mathbf{R}_{\mathbf{d}}$\;
Update $\mathbf{R}_{\mathbf{d}}=\mathbf{R}_{\mathbf{d}}(\mathbf{u},\mathbf{d},\mathcal{H})$\;
}
Update $\mathbf{R}_{\mathbf{u}}=\mathbf{R}_{\mathbf{u}}(\mathbf{u},\mathbf{d},\Delta_{\max})$\;
}
$\mathbf{u}_{m+1}\leftarrow\mathbf{u}$, $\mathbf{d}_{m+1}\leftarrow\mathbf{d}$, $\mathcal{H}_{m+1}\leftarrow\mathcal{H}$, $\Delta_{\max}^{m+1}\leftarrow\Delta_{\max}$
\end{algorithm}

\subsection{Relation to existing models}
\label{sec:rel-ex-mod}
This subsection elaborates the characteristics of existing models in order to highlight the features of the proposed framework.
There exist two kinds of PFM-based frameworks to investigate the progressive failure of fiber-reinforced composites, which differ in how the interface is treated. 

The first kind \cite{nguyen2016phase,Reddy2021,ZhangP2019,LW2020,Tarafder2020,Pengfei2020} regularizes the interface over a finite width, similar to the crack regularization in the classic PFM. Specifically, the interface is regularized with an auxiliary interface phase field $d_I:\Omega\rightarrow\mathbb{R}$ such that
$d_I-\ell_I^2 \Delta d_I =0$ in $\Omega$,  $d_I=1$ on $\Gamma_I$, and $\nabla d_I \cdot \boldsymbol{\rm{n}}=0$ on $\partial\Omega$,
where $\ell_I$ is the regularization length scale of the interface, with $\ell_I \rightarrow0$ giving the limit of the sharp interface, and $\boldsymbol{\rm{n}}$ is the outward normal vector to $\partial\Omega$. 

These contributions differ in how the interface energy is formulated. In \cite{nguyen2016phase,Reddy2021,Tarafder2020,Pengfei2020}, $\Pi_\mathrm{i}$ is transformed into the following form:
\begin{equation}
\label{reg-int-energy}
\Pi_\mathrm{i}=\int_{\Omega} \phi(\Delta,\Delta_{\max}) \gamma_I \; \mathrm{d} \Omega,
\end{equation}
to regularize the displacement jump across the interface. 
The crack surface density function of the interface is given by \cite{Shen2018-27}
$$
\gamma_I(d_I, \nabla d_I)=\frac{1}{4 c_w}\left(\frac{w\left(d_I\right)}{\ell_I}+\ell_I |\nabla d_I|^{2}\right),
$$
where $c_w=1/2$ and $w\left(d_I\right)=d_I^2$ in \cite{nguyen2016phase,Reddy2021,ZhangP2019,Tarafder2020,Pengfei2020}, and $c_w=2/3$ and $w\left(d_I\right)=d_I$ in \cite{LW2020}.
In this way, $\Delta$ is not only defined on $\Gamma_I$ but extended over the whole domain $\Omega$. This treatment may result in an extra strain component with a certain degree of arbitrariness, interfering the interplay between the elastic and plastic strains. Thus, when ductile matrix cracking is considered, the plastic strain of the matrix has to be determined from the somewhat ambiguous extra strain component as well as the total strain and the elastic strain, see Ref.~\cite{Pengfei2020}.

Somewhat differently, Refs.~\cite{ZhangP2019,LW2020} replace $\phi$ in Eq.~\eqref{reg-int-energy} by a modified interface fracture toughness to account for the influences of the fiber and the matrix on interface failure.

In the second kind \cite{paggi2017revisiting,zhang2020model,R-curve2021} and in this work, zero-thickness cohesive elements along the interface enable describing the interface failure merely by the CZM, and the interface remains sharp, eliminating the need of regularizing the interface, of solving for $d_I$, and of studying the effect of $\ell_I$ relative to other length dimensions. In \cite{paggi2017revisiting}, a linear CZM is adopted such that the critical opening displacement, playing a similar role as does $\delta_\mathrm{n}$\ here, is assumed to be dependent on the phase field $d$ in the surrounding matrix. Such dependence is realized in \cite{zhang2020model} by a different means, where an equivalent damage index is introduced, which equals the biggest among the damage index of the cohesive element of interest and those of the two adjacent bulk elements. In this work, under the spirit of \emph{minimum} ingredient, the interface debonding is assumed not to be interfered by the phase field in the bulk, which still can accurately capture the competition of the two failure mechanisms, i.e., matrix cracking and interface debonding, as will be seen in Section \ref{example}.

\section{Numerical examples}\label{example}
This section showcases the proposed framework with several numerical examples. The first example investigates the competition between crack penetration and crack deflection when a horizontal crack impinges on an inclined interface. The second example investigates cracking behaviors in fiber-reinforced composites with different strengths of the interface.

\subsection{Crack impinging on an inclined interface}
\label{sec:crack-imp}
\begin{figure}[htbp]
	\centering
	\includegraphics[width=0.8\textwidth]{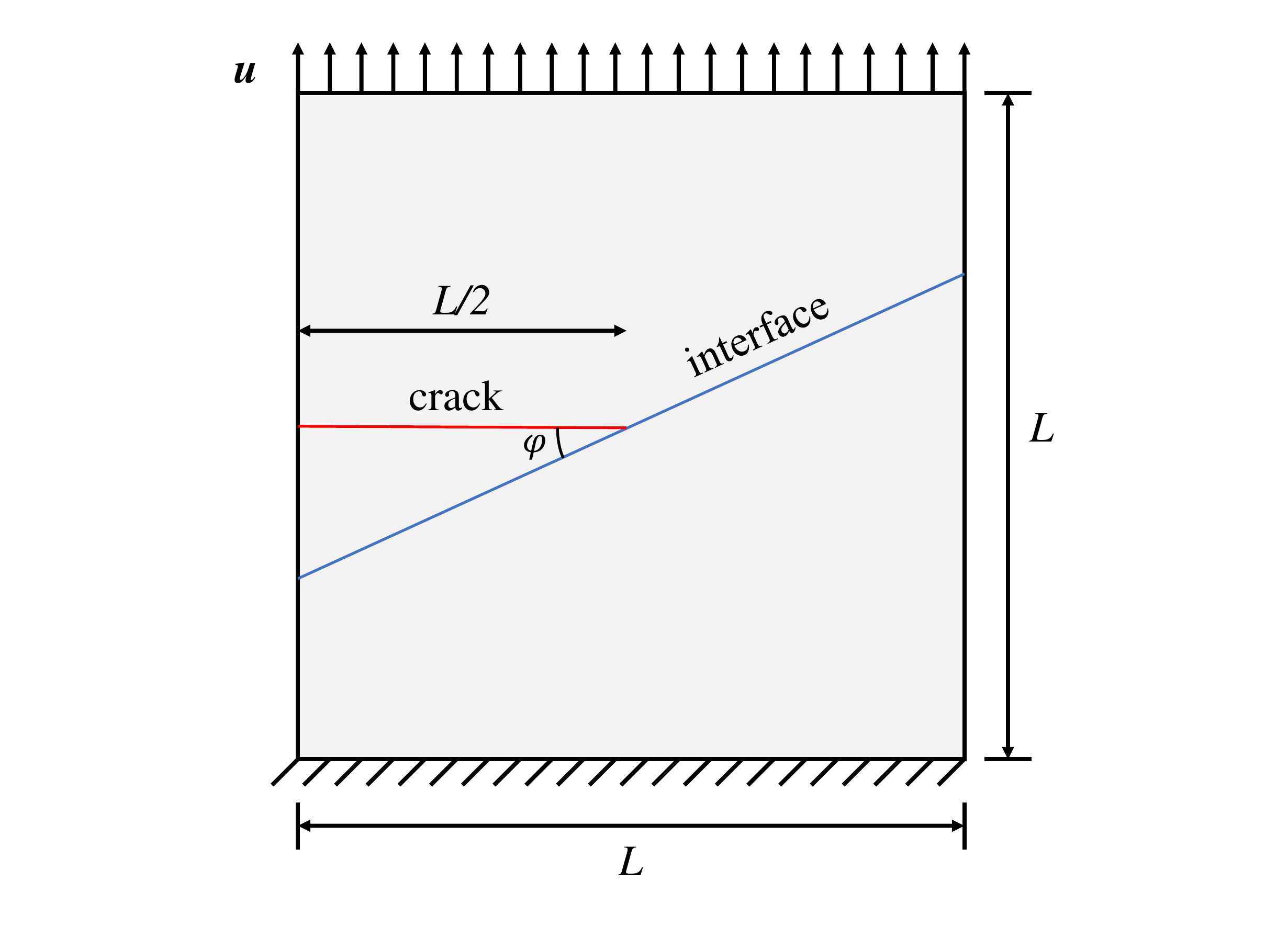}
	\caption{Geometry and boundary conditions of a square plate with an initial crack impinging on an inclined interface with $L=10$ mm and $u=0.09$ mm; $\varphi\in[0,90^\circ]$ for different cases.}
	\label{case1}
\end{figure}

\begin{figure}[htbp]
    \centering
    \includegraphics[width=0.8\textwidth]{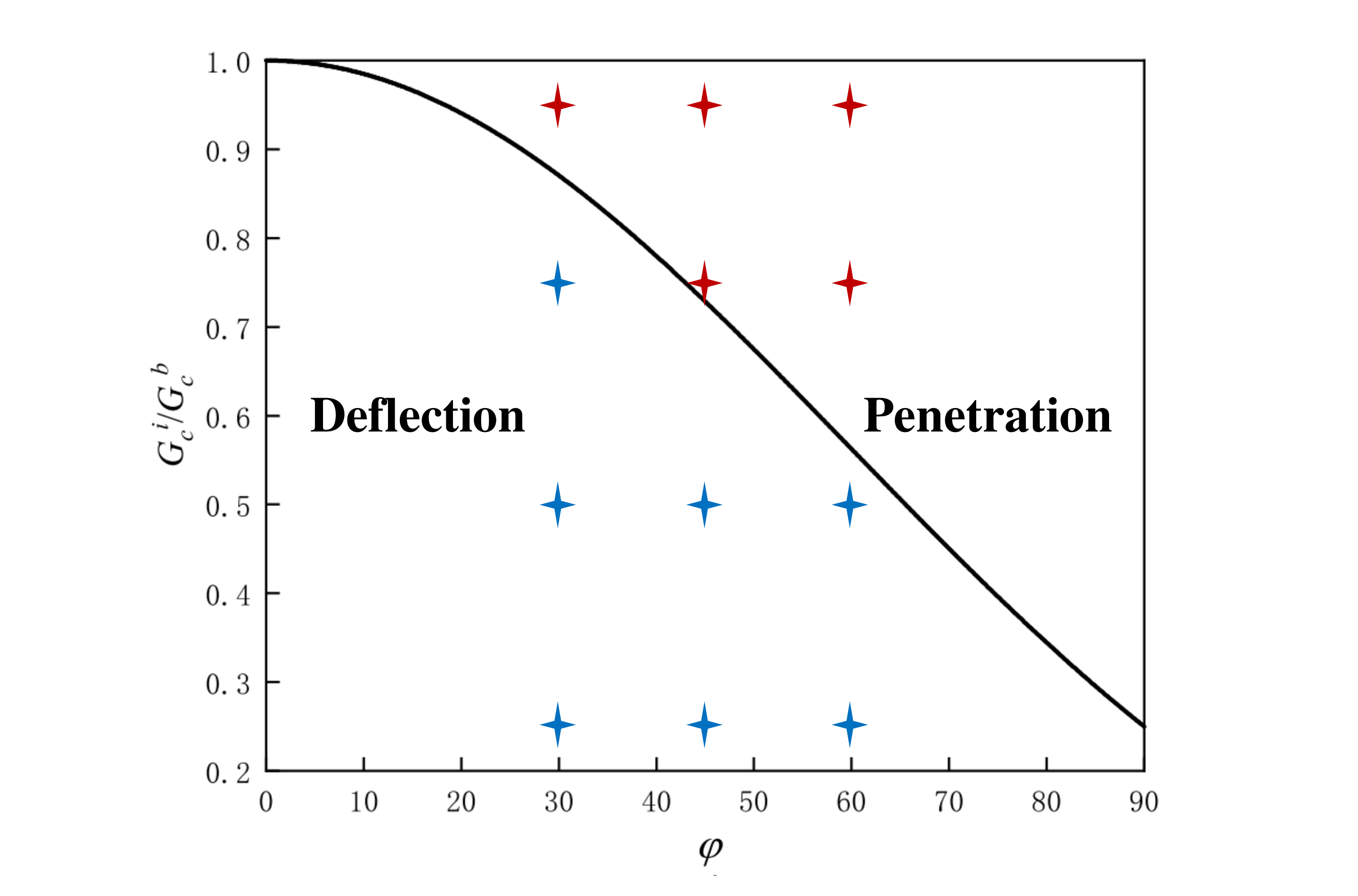}
    \caption{The deflection-penetration curve obtained according to Eq.~\eqref{analytical}. Overlaid are markers representing the numerical results performed in Section \ref{sec:crack-imp}, where deflection cases are shown in blue and penetration cases in red. These results are consistent with the curve.}
    \label{penetration-deflection}
\end{figure}

In this subsection, when a horizontal crack impinges on an inclined interface, we examine the competition between the crack propagation modes (penetration versus deflection). As illustrated in Fig.~\ref{case1}, the crack encounters the interface with the inclination angle $\varphi\in[0,90^\circ]$ in a square domain under uniaxial tension. The crack may either penetrate into the bulk or deflect along the interface, for which an analytical solution was obtained by He and Hutchinson \cite{ming1989crack} in which the crack propagation mode depends on the relative size of (a) the ratio of $G_{c}^{i}$, the critical energy release rate of the interface, and $G_{c}^{b}$, the critical energy release rate of the bulk material and (b) the following ratio:
\begin{equation}\label{analytical}
  \eta=\frac{G_c^d}{G_c^p}=\frac{1}{16}\left\{\left[3 \cos \left(\frac{\varphi}{2}\right)+\cos \left( \frac{3\varphi}{2}\right)\right]^{2}+\left[\sin \left(\frac{\varphi}{2}\right)+\sin \left( \frac{3\varphi}{2}\right)\right]^{2}\right\},
\end{equation}
where $G_{c}^{d}$ is the critical energy release rate for crack deflection along the interface and $G_{c}^{p}$ is the critical energy release rate for crack penetration into the bulk. If
$$
\frac{G_{c}^{i}}{G_{c}^{b}}<\eta,
$$
the crack is deflected, otherwise the crack penetrates into the bulk. A curve based on Eq.~\eqref{analytical} separating deflection and penetration is shown in Fig.~\ref{penetration-deflection}.

Particularly, in this numerical example, the geometric parameters under consideration are $L=10$ mm and $u=0.09$ mm with a uniform increment $\Delta u=0.005$ mm. The material proprieties adopted are: Young's modulus $E=210$ GPa, Poisson's ratio $\nu=0.3$, $\sigma_{\mathrm{max}}=600$ MPa, $G_{c}^{i}=0.32$ N/mm, and $\ell=0.1$ mm.

\begin{figure}[htbp]
\centering
	\subfigure[$G_{c}^{i}/G_{c}^{b}=0.25$: deflection]{
	\centering
	\includegraphics[width=0.4\textwidth]{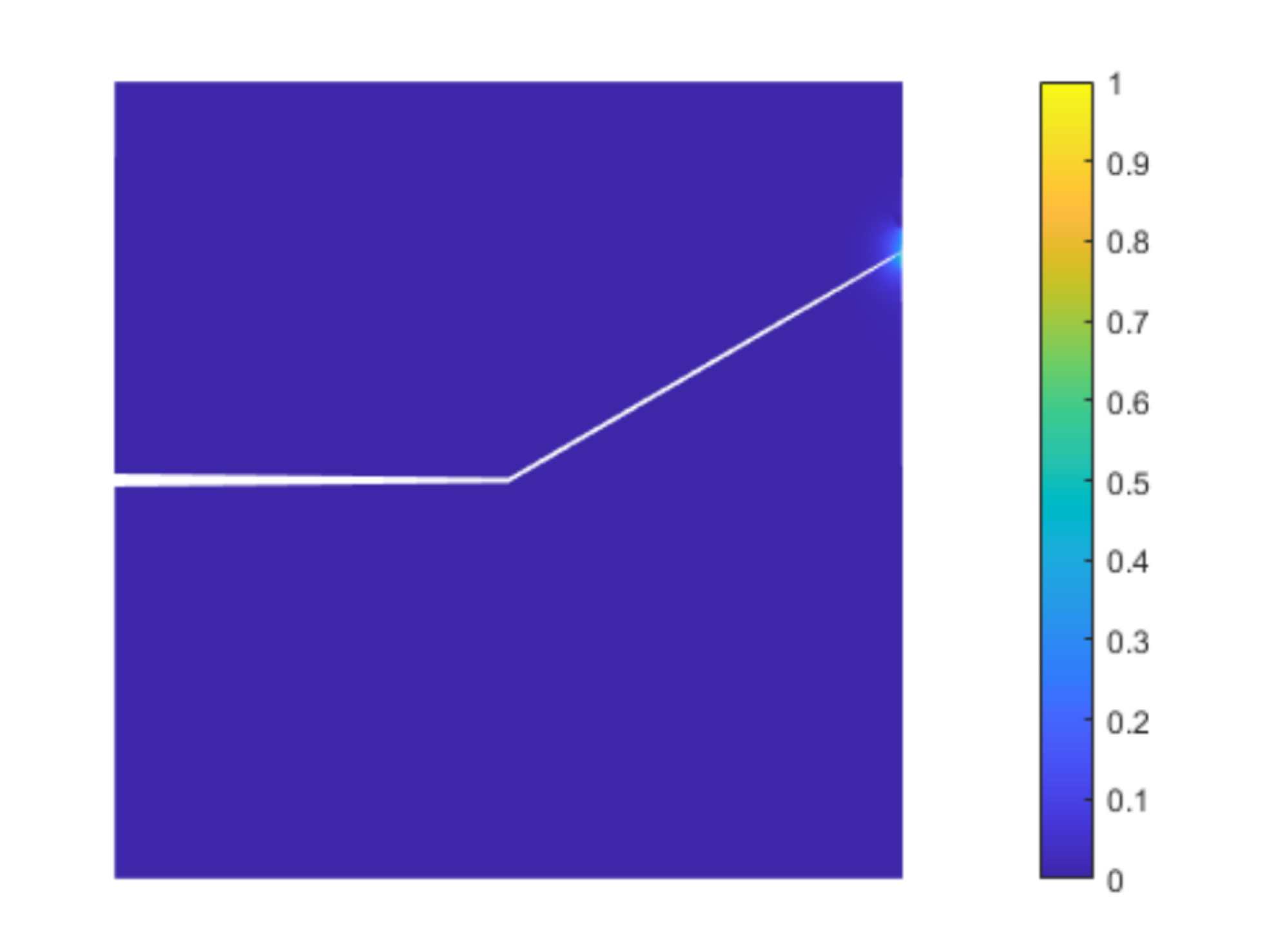}
	}
	\subfigure[$G_{c}^{i}/G_{c}^{b}=0.5$: deflection]{
	\centering
	\includegraphics[width=0.4\textwidth]{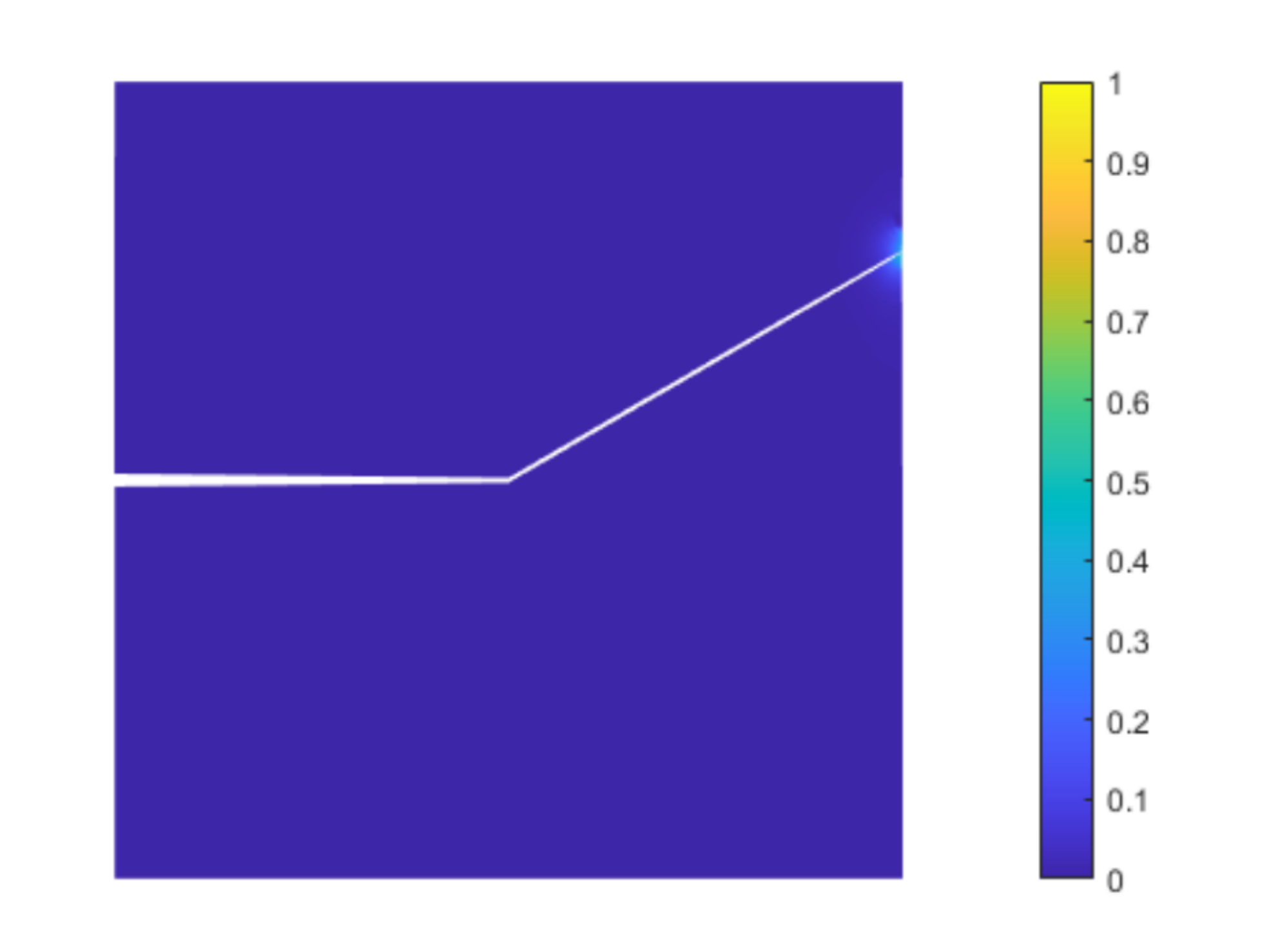}
	}
	\subfigure[$G_{c}^{i}/G_{c}^{b}=0.75$: deflection]{
	\centering
	\includegraphics[width=0.4\textwidth]{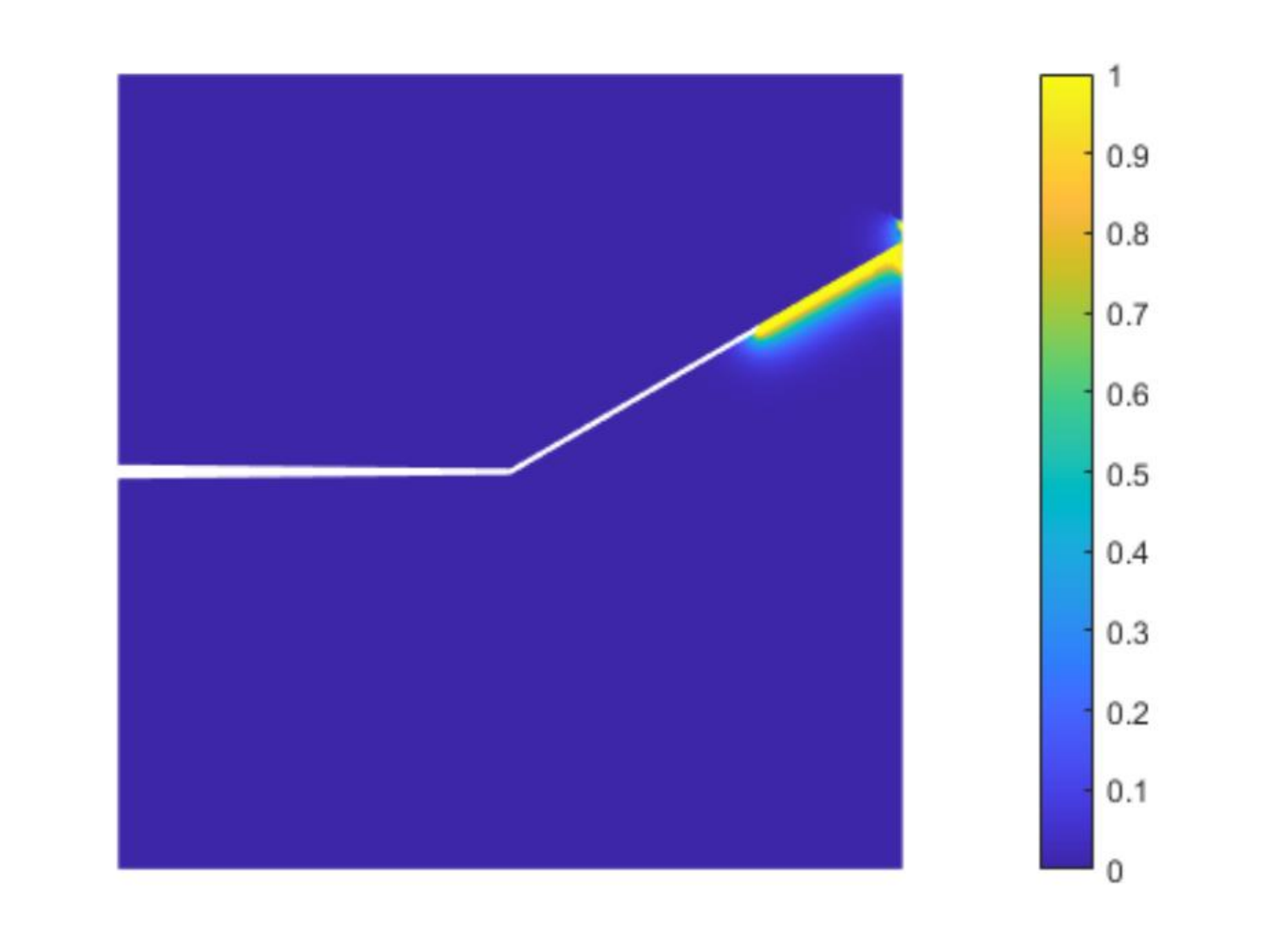}
	}
	\subfigure[$G_{c}^{i}/G_{c}^{b}=0.95$: penetration]{
	\centering
	\includegraphics[width=0.4\textwidth]{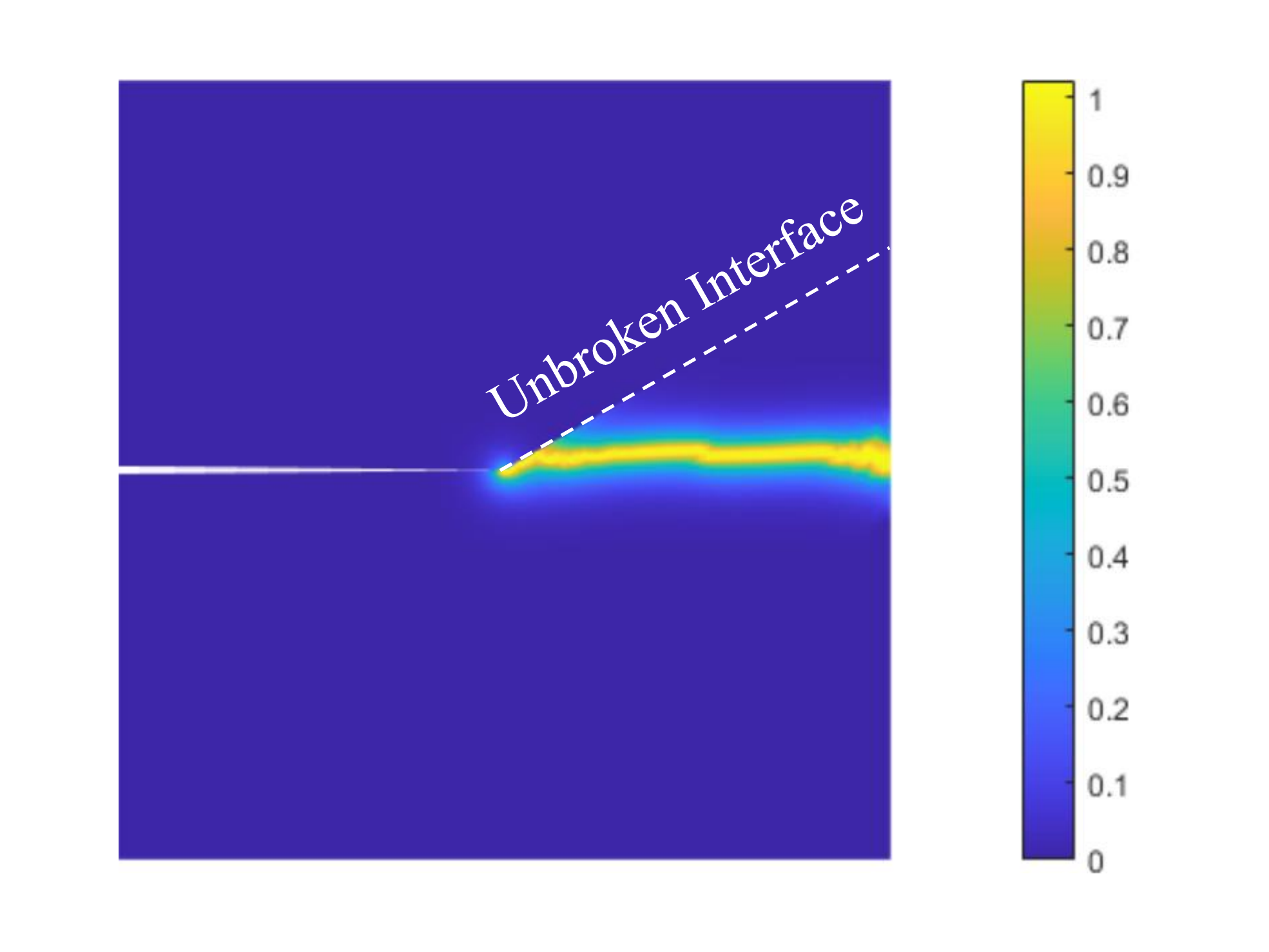}
	}
\caption{Phase field result of the crack impinging test: (a) $G_{c}^{i}/G_{c}^{b}=0.25$, a case of deflection, (b) $G_{c}^{i}/G_{c}^{b}=0.5$, a case of deflection, (c) $G_{c}^{i}/G_{c}^{b}=0.75$, a case of deflection, and (d) $G_{c}^{i}/G_{c}^{b}=0.95$, a case of penetration, for $\varphi=30^{\circ}$.}
\label{angle-30}
\end{figure}

\begin{figure}[htbp]
\centering
	\subfigure[$G_{c}^{i}/G_{c}^{b}=0.25$: deflection]{
	\centering
	\includegraphics[width=0.4\textwidth]{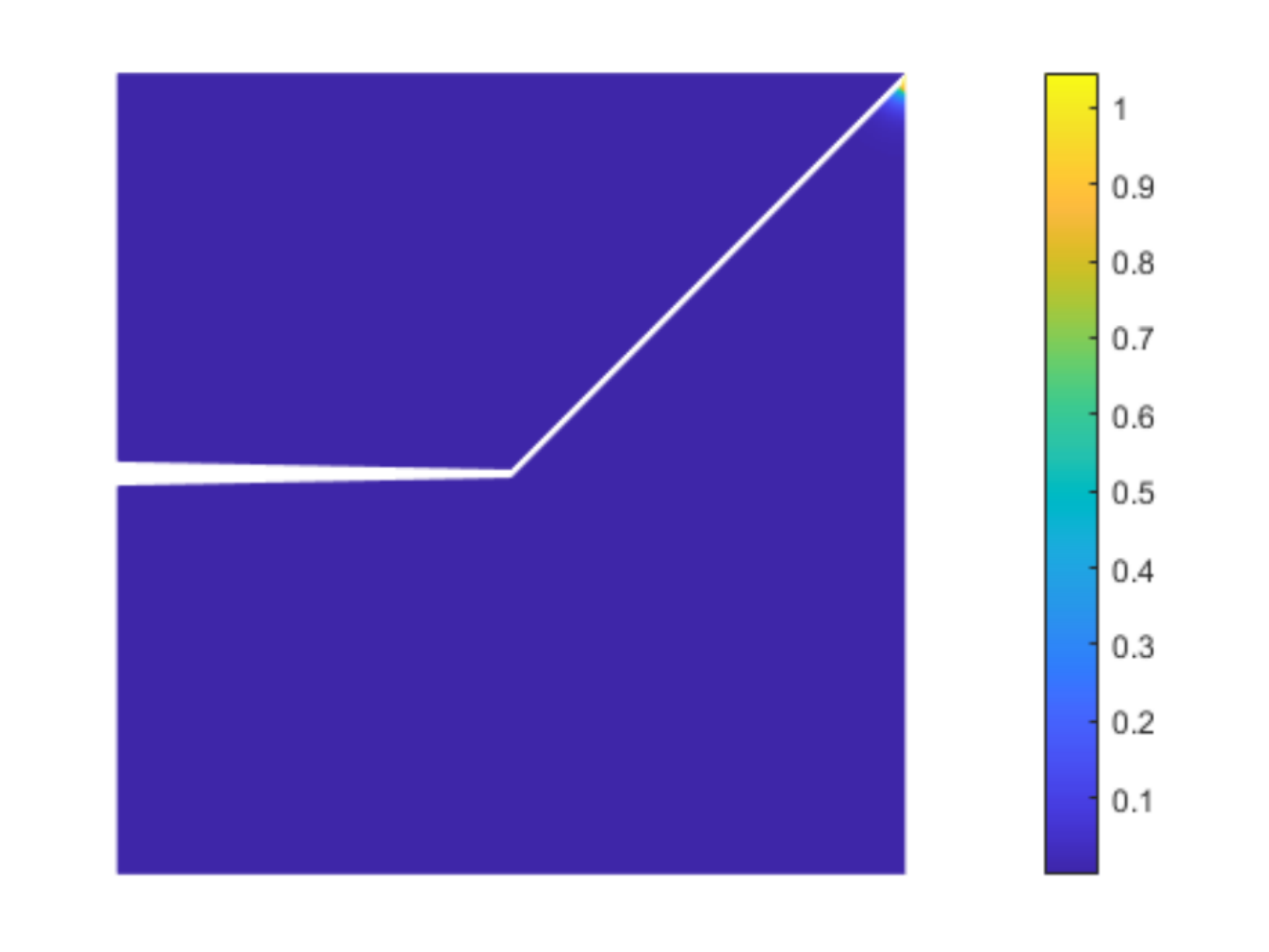}
	}
	\subfigure[$G_{c}^{i}/G_{c}^{b}=0.5$: deflection]{
	\centering
	\includegraphics[width=0.4\textwidth]{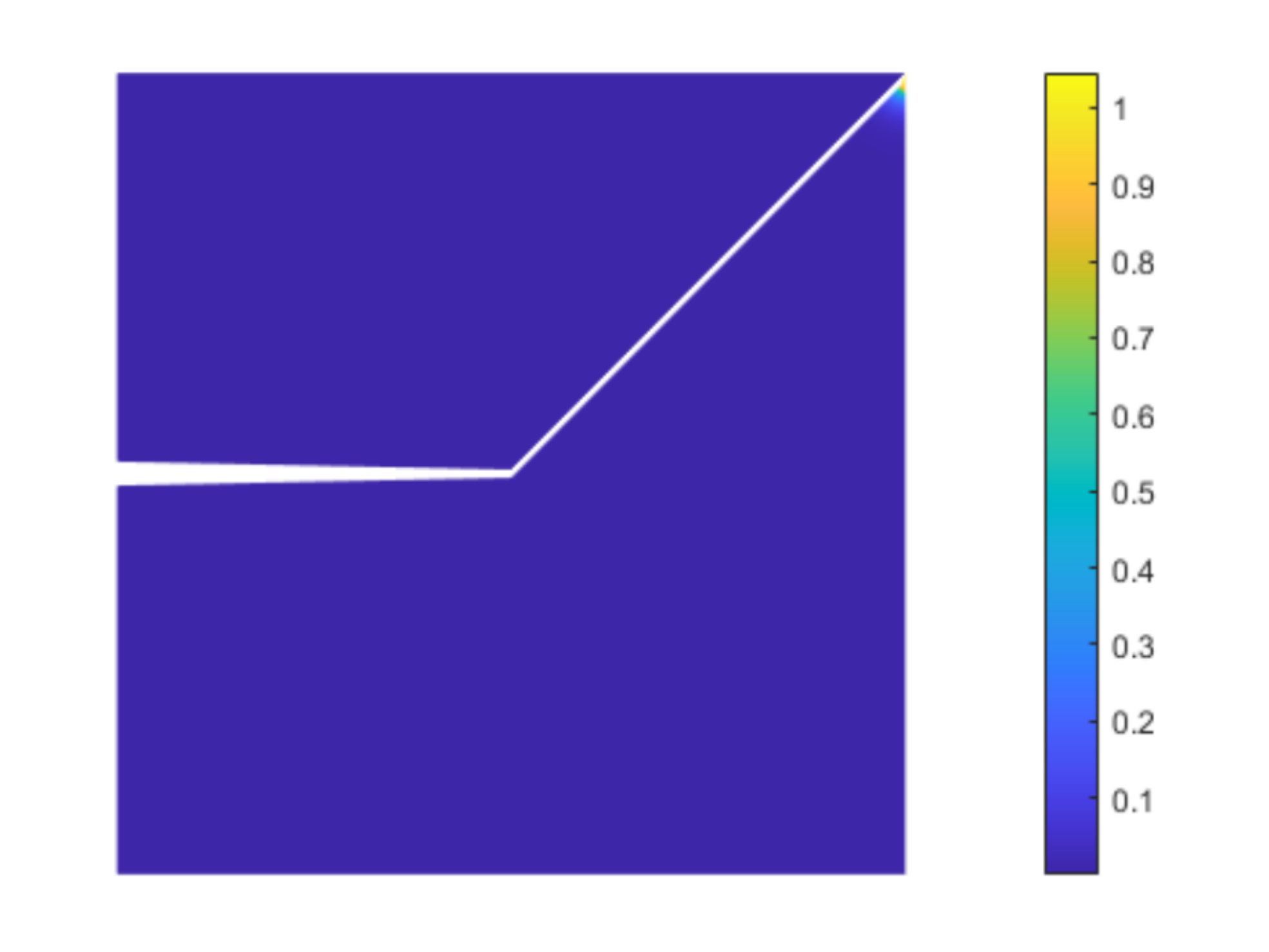}
	}
	\subfigure[$G_{c}^{i}/G_{c}^{b}=0.75$: deflection]{
	\centering
	\includegraphics[width=0.4\textwidth]{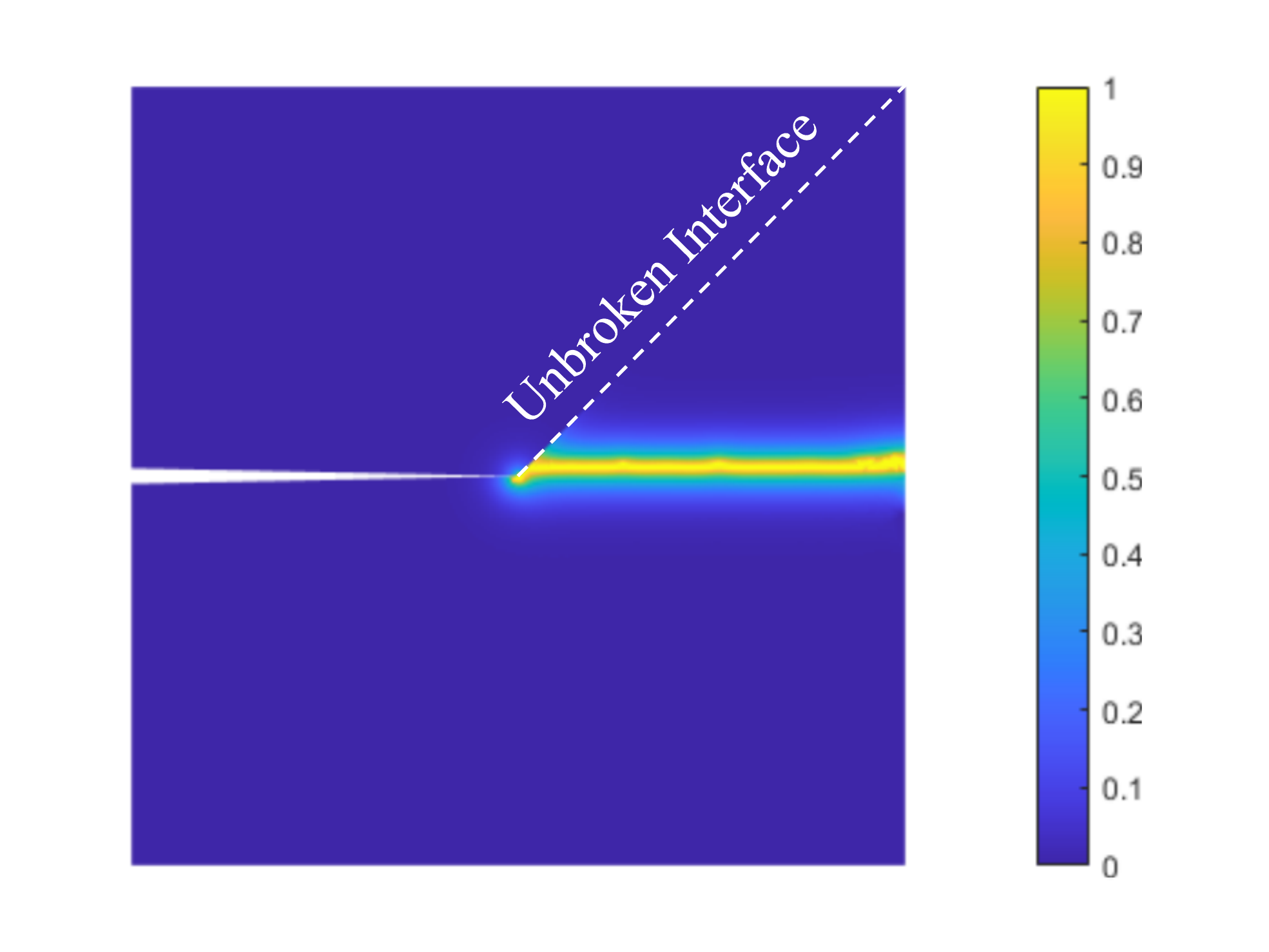}
	}
	\subfigure[$G_{c}^{i}/G_{c}^{b}=0.95$: penetration]{
	\centering
	\includegraphics[width=0.4\textwidth]{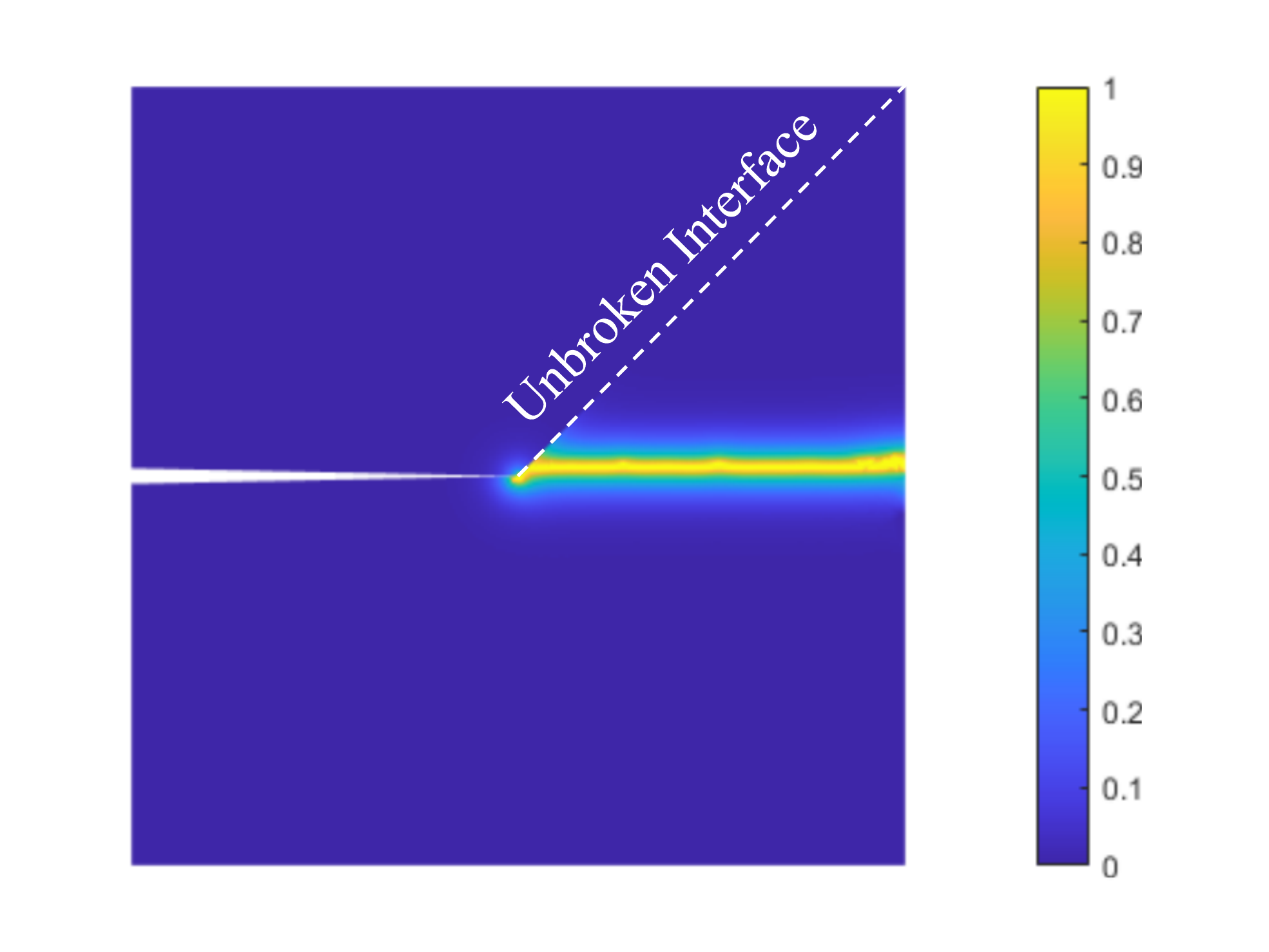}
	}
\caption{Phase field result of the crack impinging test: (a) $G_{c}^{i}/G_{c}^{b}=0.25$, a case of deflection, (b) $G_{c}^{i}/G_{c}^{b}=0.5$, a case of deflection, (c) $G_{c}^{i}/G_{c}^{b}=0.75$, a case of penetration, and (d) $G_{c}^{i}/G_{c}^{b}=0.95$, a case of penetration, for $\varphi=45^{\circ}$.}
\label{angle-45}
\end{figure}

\begin{figure}[htbp]
\centering
	\subfigure[$G_{c}^{i}/G_{c}^{b}=0.25$: deflection]{
	\centering
	\includegraphics[width=0.4\textwidth]{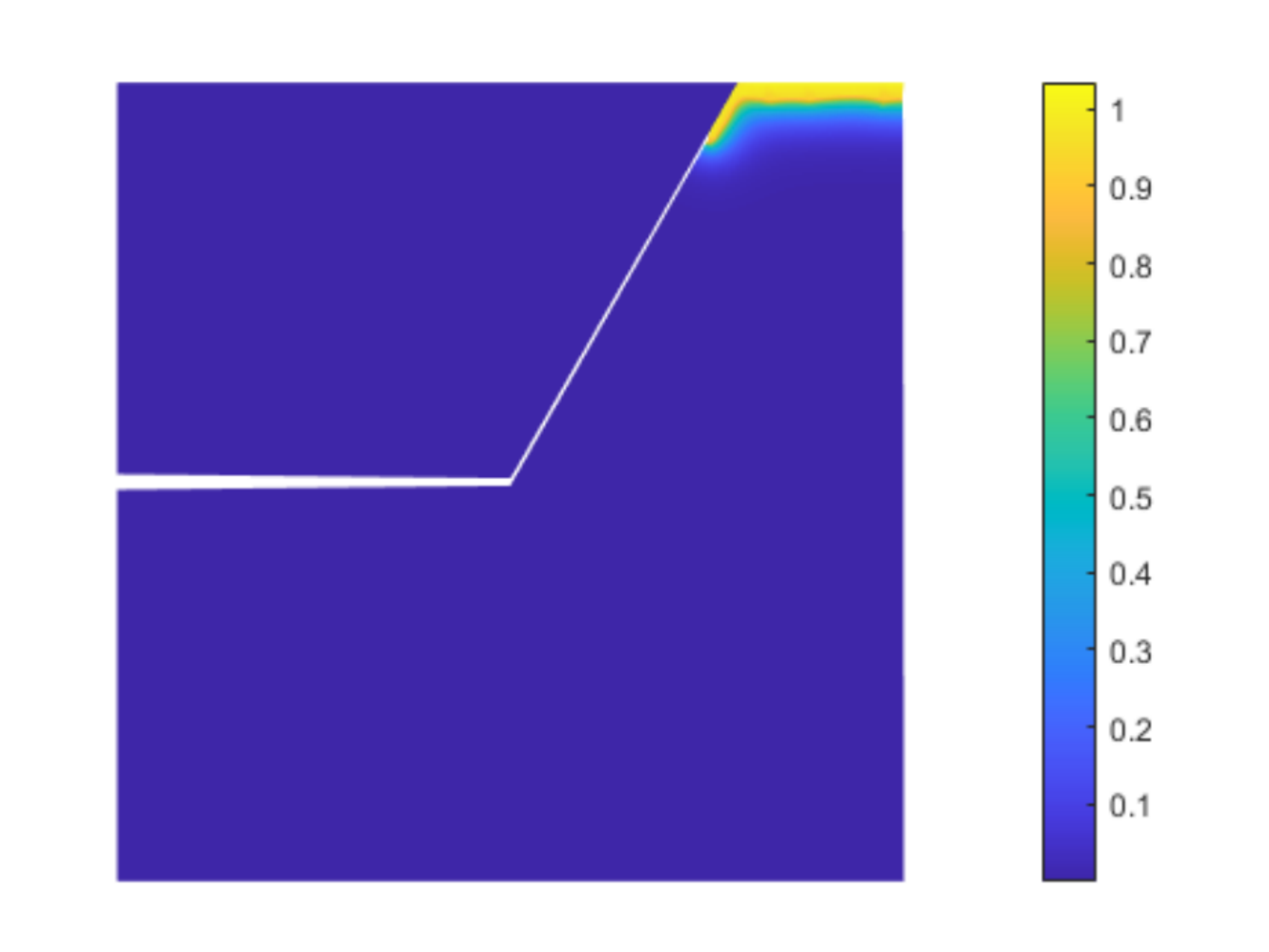}
	}
	\subfigure[$G_{c}^{i}/G_{c}^{b}=0.5$: deflection]{
	\centering
	\includegraphics[width=0.4\textwidth]{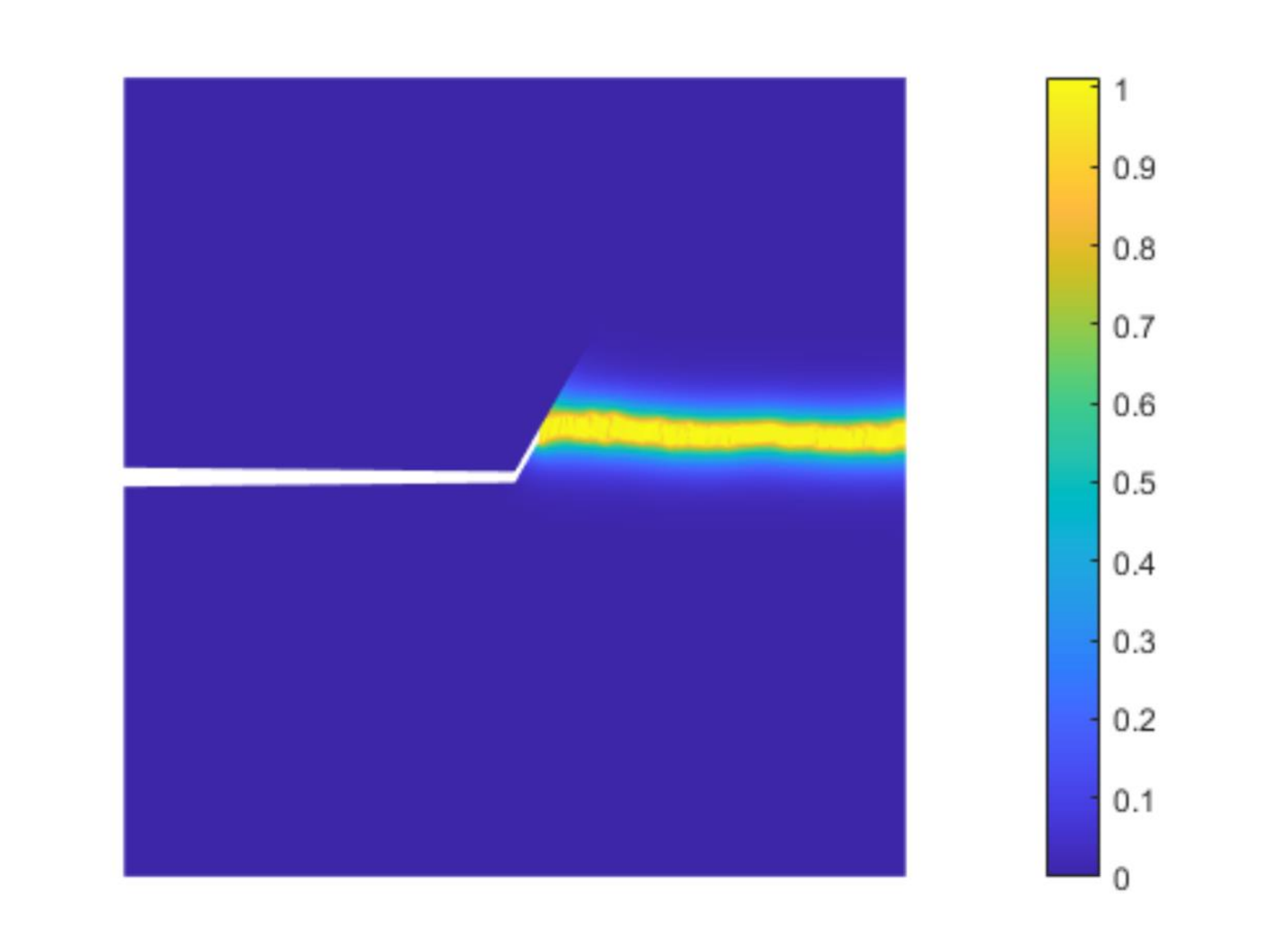}
	}
	\subfigure[$G_{c}^{i}/G_{c}^{b}=0.75$: penetration]{
	\centering
	\includegraphics[width=0.4\textwidth]{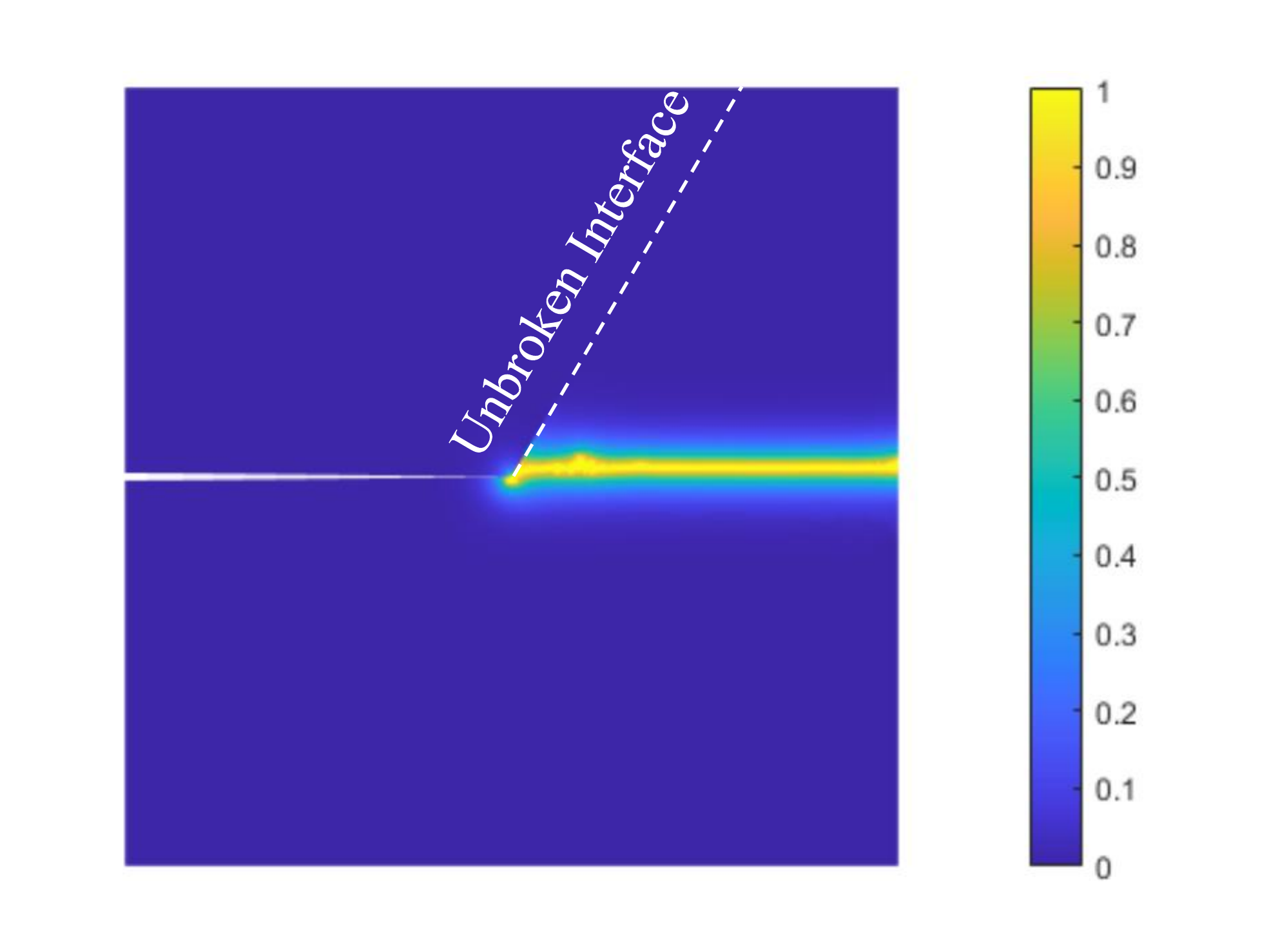}
	}
	\subfigure[$G_{c}^{i}/G_{c}^{b}=0.95$: penetration]{
	\centering
	\includegraphics[width=0.4\textwidth]{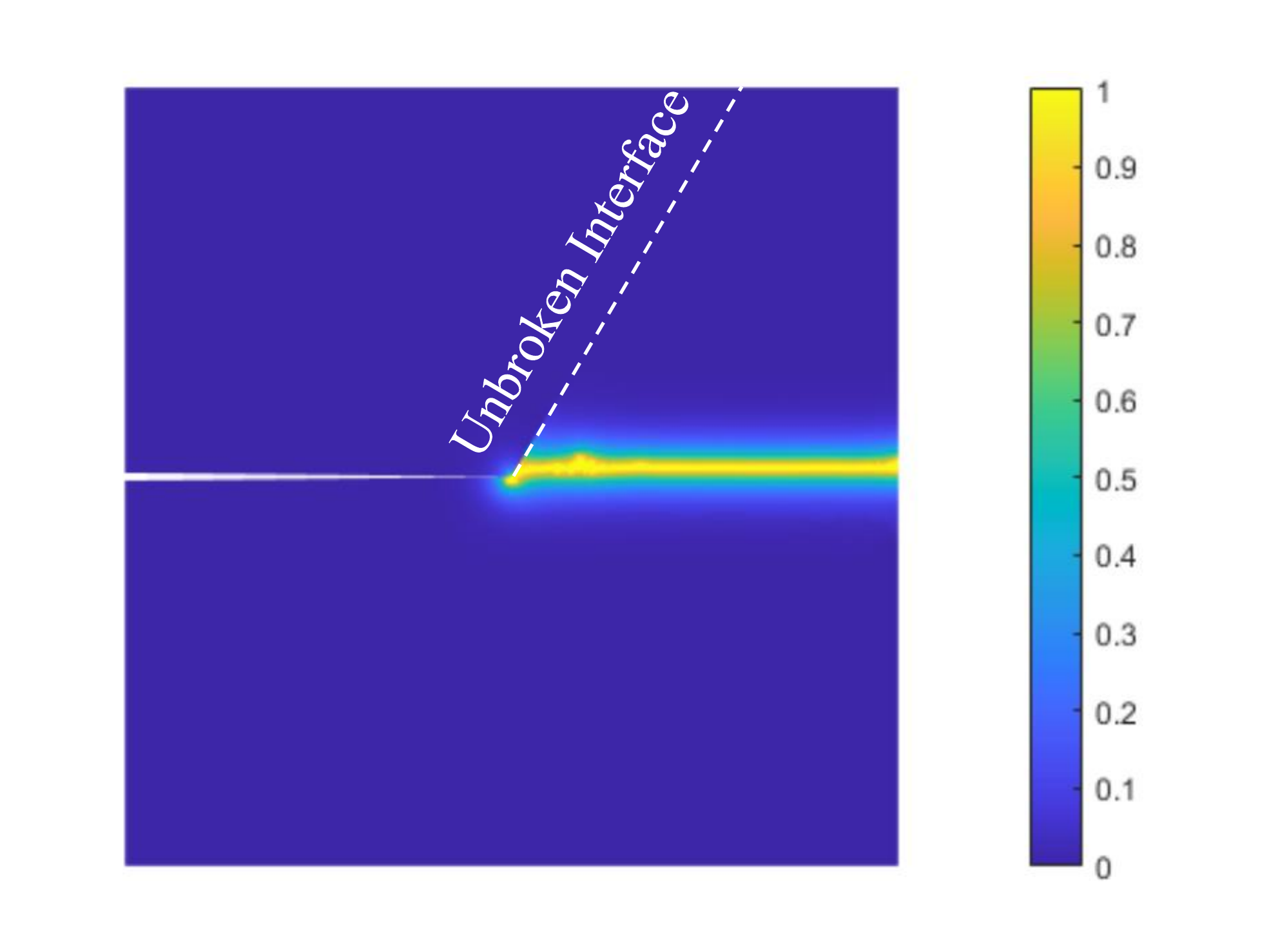}
	}
\caption{Phase field result of the crack impinging test: (a) $G_{c}^{i}/G_{c}^{b}=0.25$, a case of deflection, (b) $G_{c}^{i}/G_{c}^{b}=0.5$, a case of deflection, (c) $G_{c}^{i}/G_{c}^{b}=0.75$, a case of penetration, and (d) $G_{c}^{i}/G_{c}^{b}=0.95$, a case of penetration, for $\varphi=60^{\circ}$.}
\label{angle-60}
\end{figure}

The crack profiles for deflection or penetration of a total of 12 cases, i.e., $\varphi\in\{30^{\circ}, 45^{\circ}, 60^{\circ}\}$ under $G_{c}^{i}/G_{c}^{b}\in\{0.25, 0.5, 0.75, 0.95\}$, are compared with the said analytical results in Figs.~\ref{angle-30}, \ref{angle-45}, and \ref{angle-60}. It can be clearly seen that even with the \emph{minimum} ingredient, the proposed framework is capable of accurately capturing the competition between crack penetration and crack deflection, or in other words, the competition between matrix cracking and interface debonding.

\subsection{Cracking behaviors in fiber-reinforced composites}

In this subsection, we investigate the cracking behaviors of a fiber-matrix system. Firstly, in Section \ref{brittle-pre-crack}, we study the case of the system with a pre-existing crack under tension under different strengths of the interface, where the matrix is set to be brittle. Secondly, in Section \ref{ductile-pre-crack}, we take into account the elastoplasticity of the matrix.

\subsubsection{Brittle matrix cracking and interface debonding in fiber-reinforced composites}
\label{brittle-pre-crack}

\begin{figure}[htbp]
	\centering
	\includegraphics[width=0.8\textwidth]{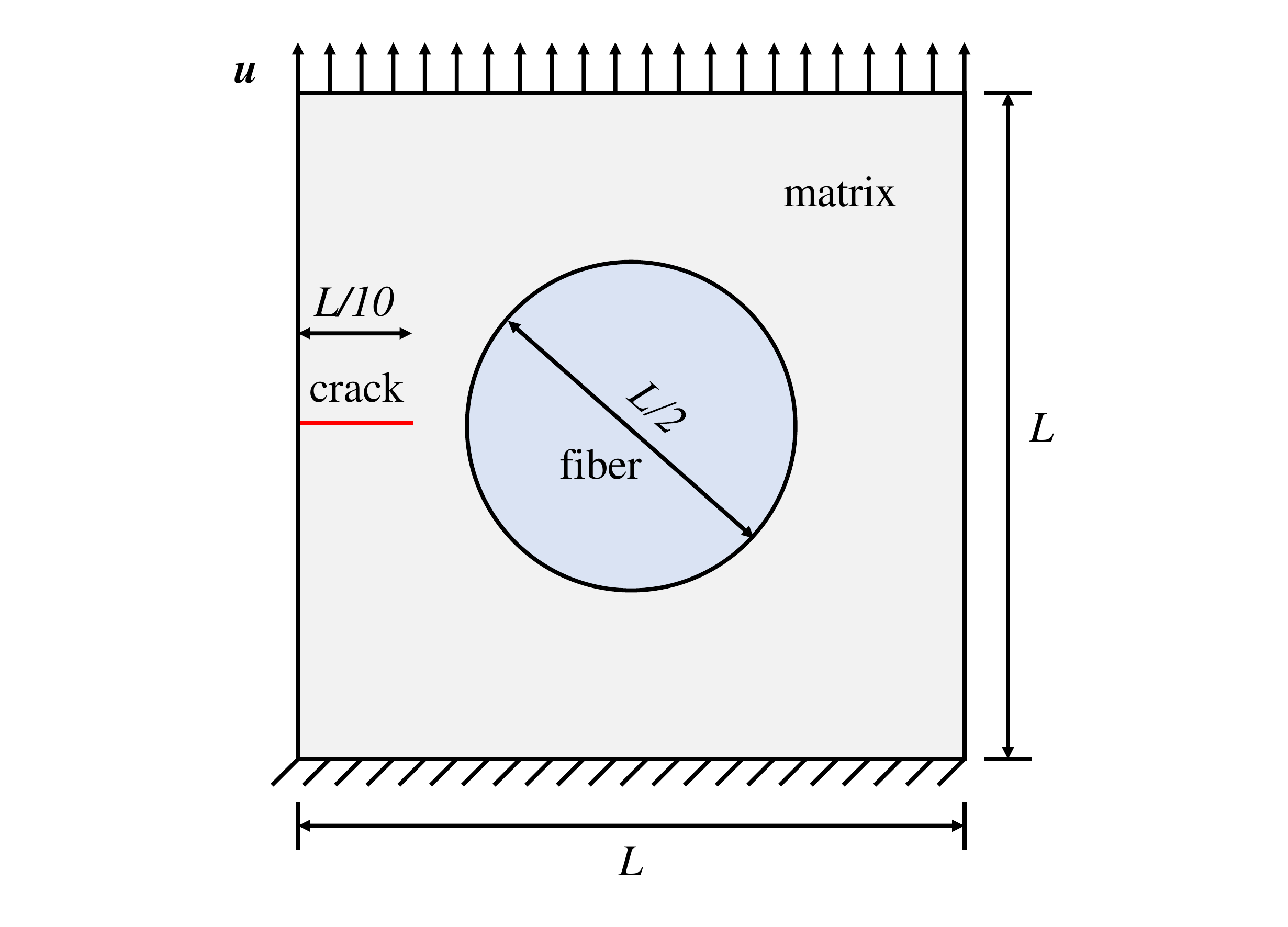}
	\caption{Geometry and boundary conditions of a square fiber-reinforced composite plate with an initial crack. Here $L=10$ mm and $u=0.15$ mm.}
	\label{case2}
\end{figure}

In this numerical example, we consider a brittle matrix, i.e., $\sigma_\mathrm{Y}^0=\infty$. A fiber-matrix system under uniaxial tension with an initial crack is shown in Fig.~\ref{case2}, in whih $L=10$ mm and $u=0.15$ mm with an equal-sized increment $\Delta u=0.005$ mm. The material properties of the fiber are: Young's modulus $E_f=413$ GPa, Poisson's ratio $\nu_f=0.33$, energy release rate $G_{c}^{f}=43$ N/mm; the material properties of the matrix are: Young's modulus $E_m=125$ GPa, Poisson's ratio $\nu_m=0.3$, energy release rate $G_{c}^{m}=20$ N/mm; for both the fiber and the matrix, $\ell=0.3$ mm. We consider two strengths for the interface: a stronger one with a critical energy release rate $G_{c}^{i}=50$ N/mm and a weaker one with $G_{c}^{i}=3$ N/mm; $\delta_n$ for both kinds of interfaces is set to 0.01 mm. 

\begin{figure}[htbp]
\centering
	\subfigure[Stronger interface: $G_{c}^{i}=50$ N/mm]{
	\centering
	\includegraphics[width=0.5\textwidth]{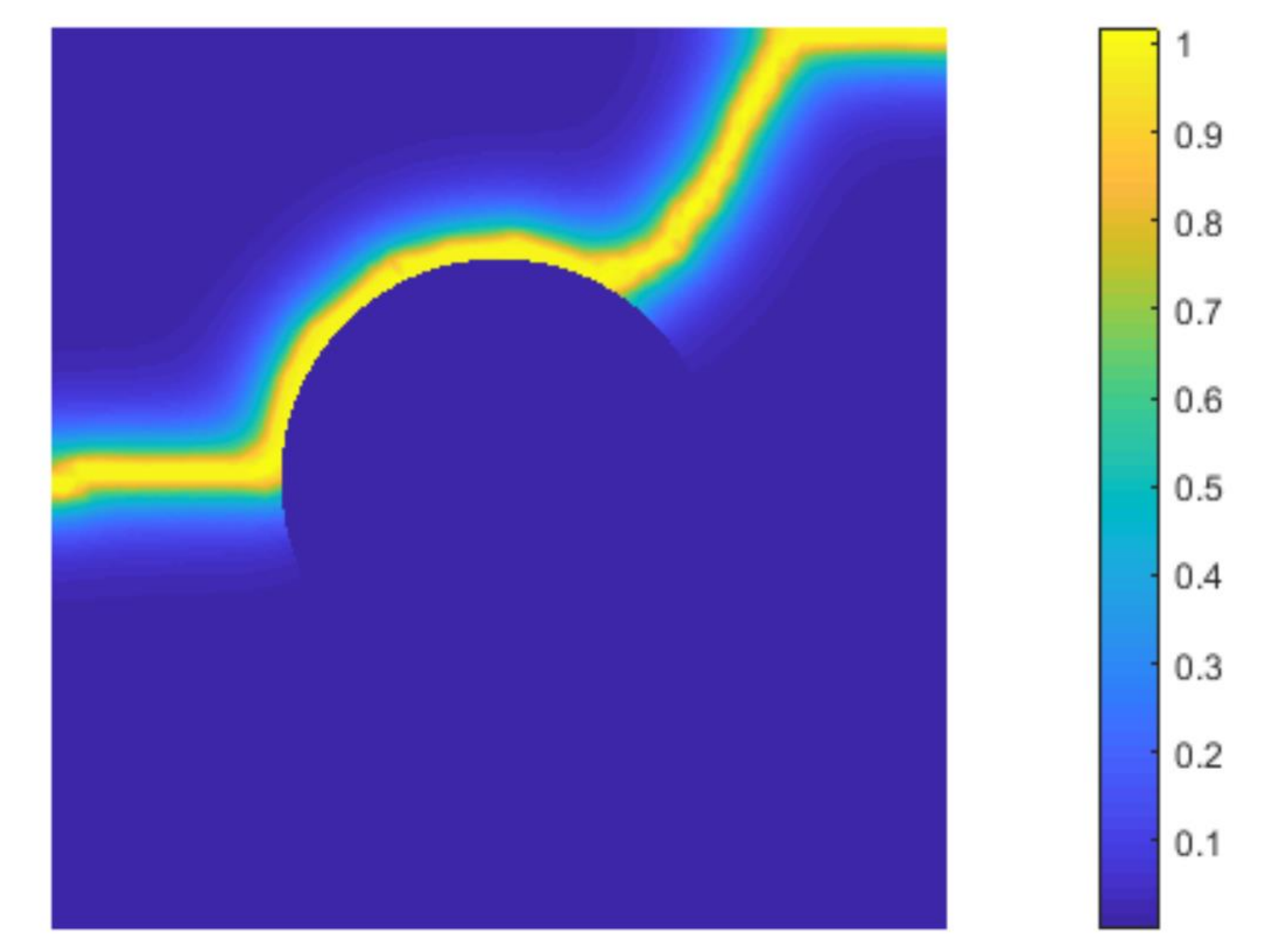}
	}
	\subfigure[Weaker interface: $G_{c}^{i}=3$ N/mm]{
	\centering
	\includegraphics[width=0.5\textwidth]{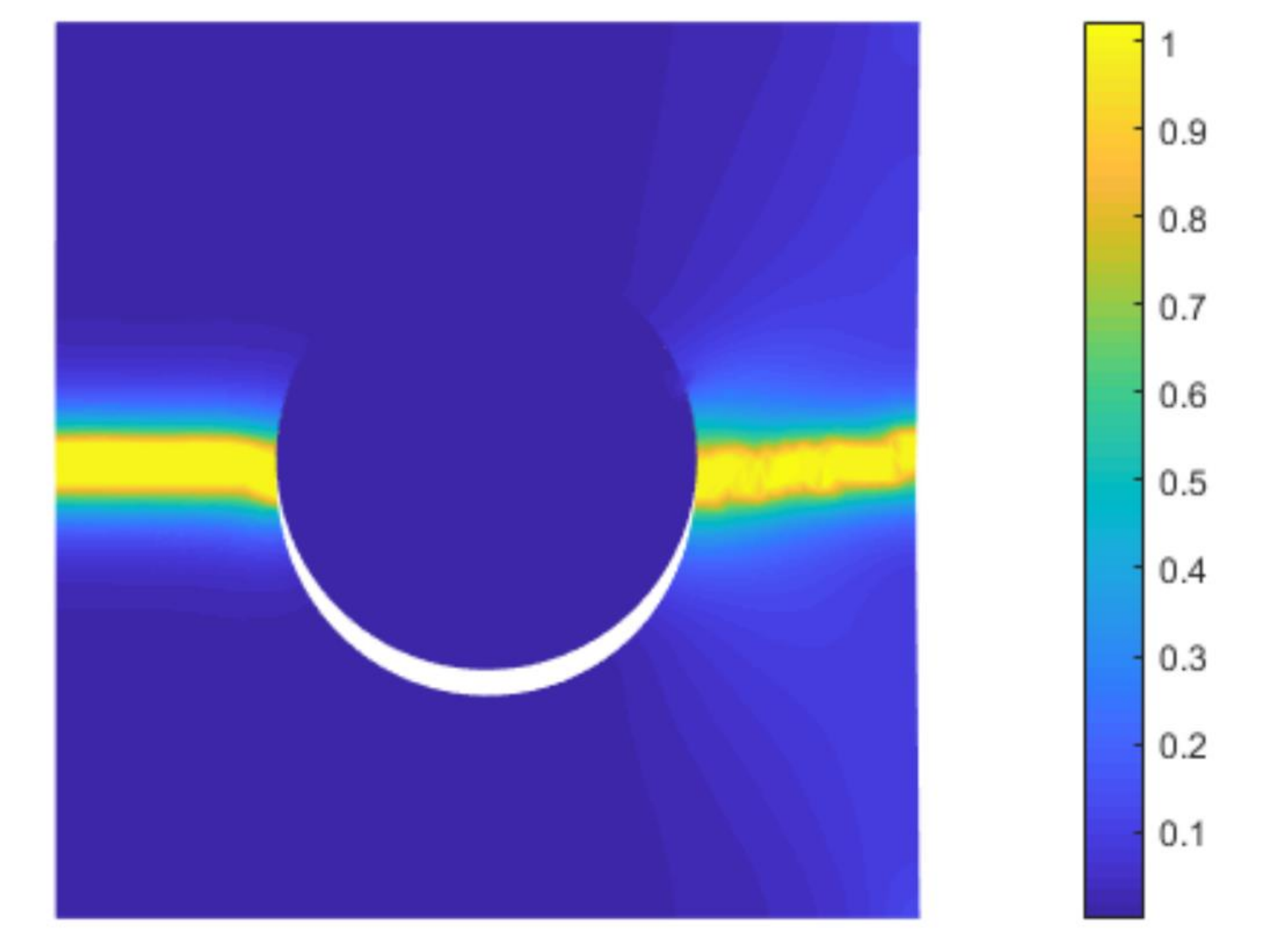}
	}
\caption{Crack propagation patterns in terms of phase field profiles in a composite with possible brittle matrix cracking and interface debonding for (a) a stronger interface, and (b) a weaker interface.}
\label{strong-weak}
\end{figure}

The crack profiles after loading are shown in Fig.~\ref{strong-weak}. For the weaker interface, we find that both matrix cracking and interface debonding occur. In contrast, for the stronger interface, the matrix crack impinges on the interface and further propagates along the interface without interface debonding.

\subsubsection{Ductile matrix cracking and interface debonding in fiber-reinforced composites}
\label{ductile-pre-crack}

Herein, we consider an elastoplastic matrix  by setting $\sigma_\mathrm{Y}^0=0.3$ GPa and $K=6$ GPa. The geometric information, boundary conditions, and the other material properties remain the same as those in Section \ref{brittle-pre-crack}. 

\begin{figure}[htbp]
\centering
	\subfigure[Stronger interface: $G_{c}^{i}=50$N/mm]{
	\centering
	\includegraphics[width=0.5\textwidth]{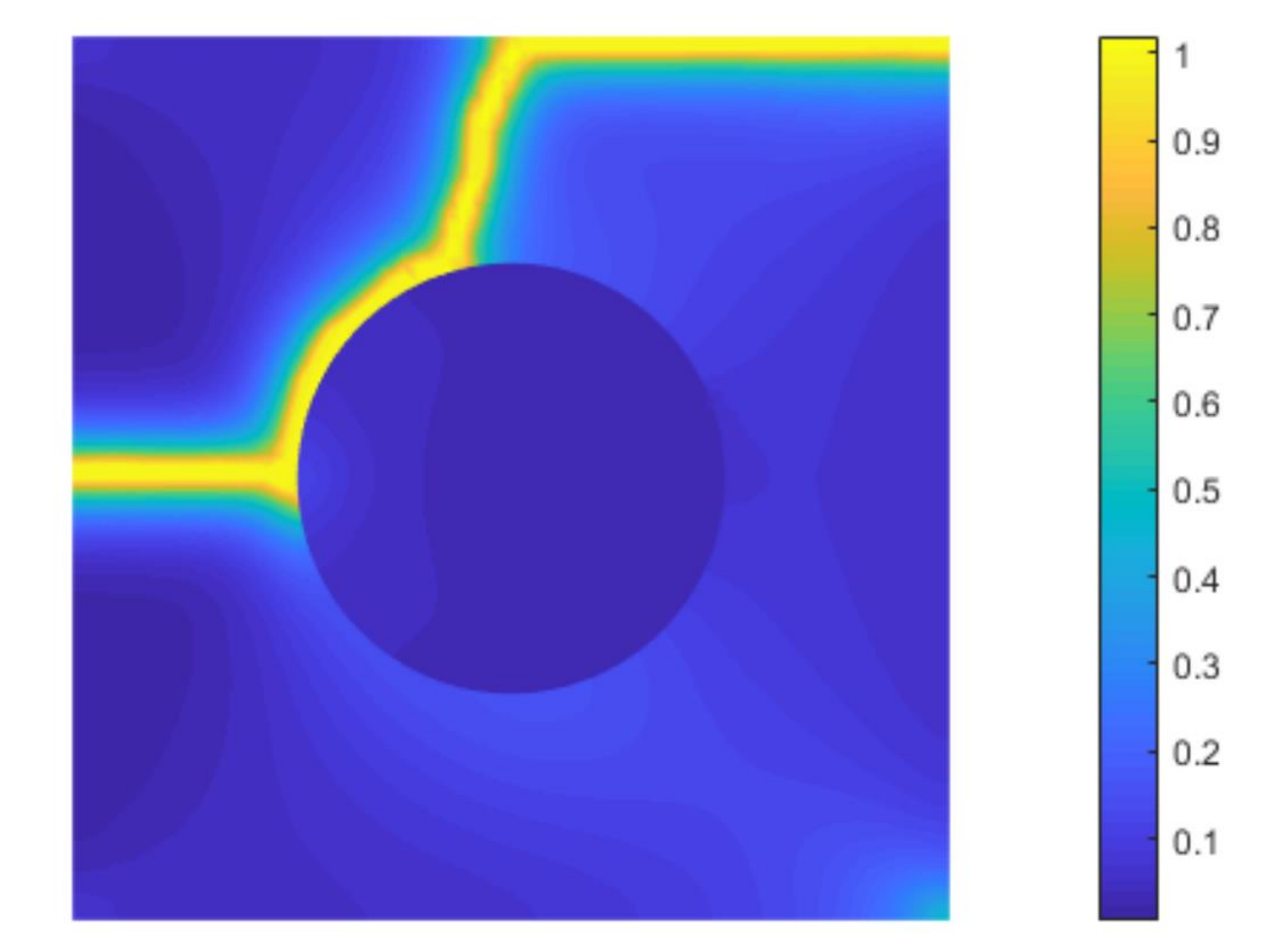}
	}
	\subfigure[Weaker interface: $G_{c}^{i}=3$N/mm]{
	\centering
	\includegraphics[width=0.5\textwidth]{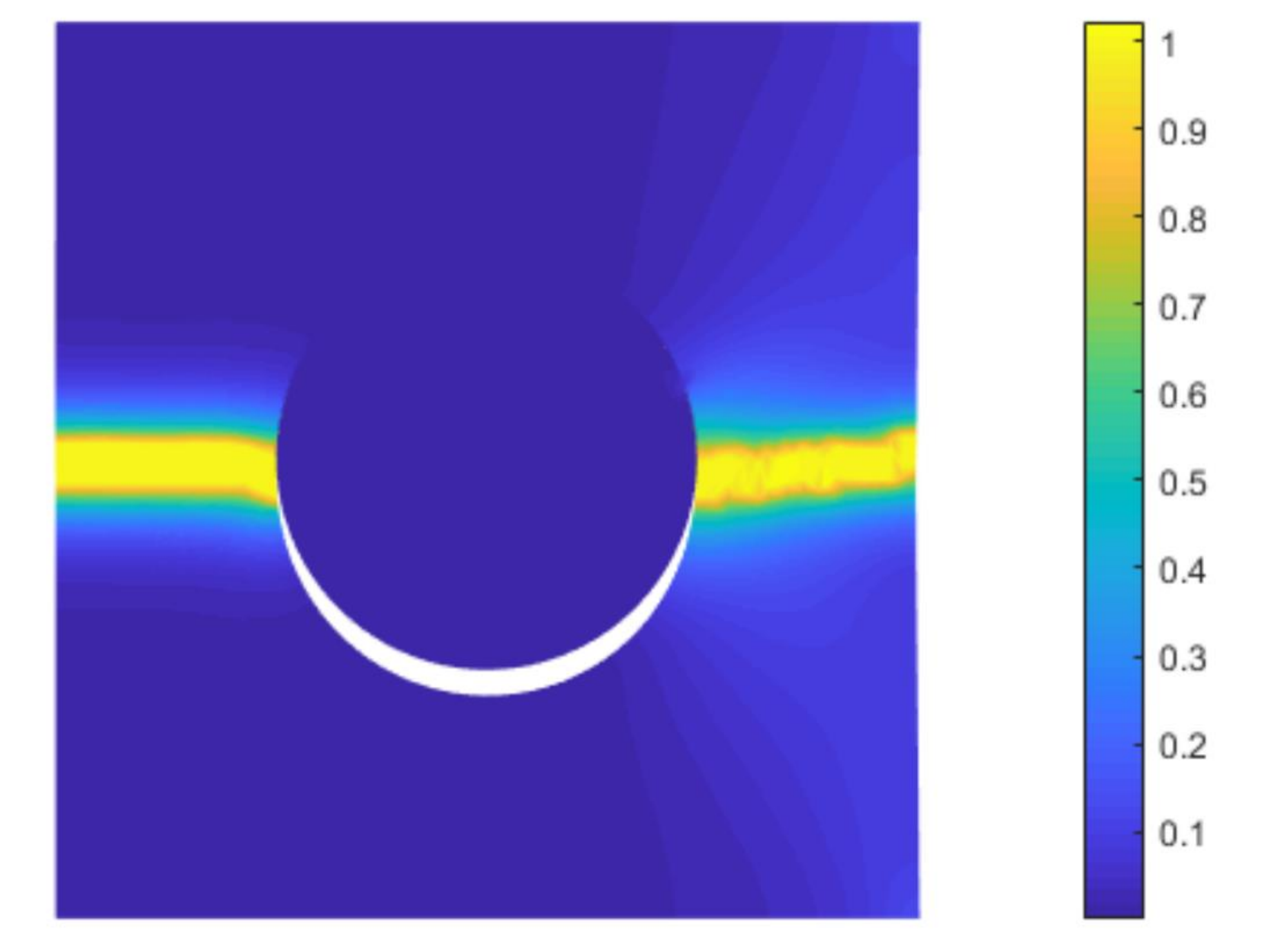}
	}
\caption{Crack propagation patterns and corresponding phase field profiles in a composite with possible ductile matrix cracking and interface debonding for (a) stronger interface, and (b) weaker interface.}
\label{strong-weak-ductile}
\end{figure}

\begin{figure}[htbp]
\centering
	\subfigure[Stronger interface: $G_{c}^{i}=50$N/mm]{
	\centering
	\includegraphics[width=0.5\textwidth]{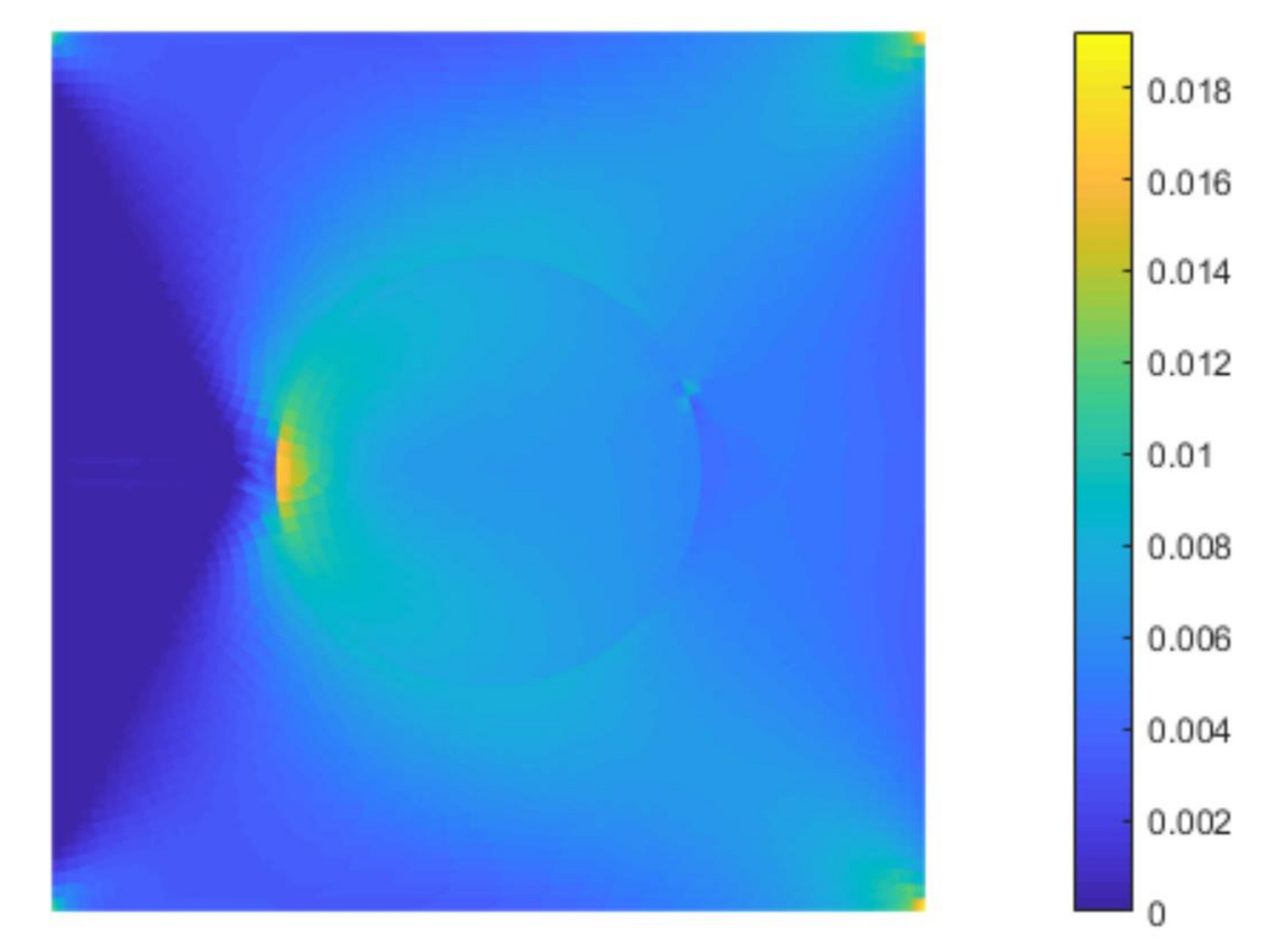}
	}
	\subfigure[Weaker interface: $G_{c}^{i}=3$N/mm]{
	\centering
	\includegraphics[width=0.5\textwidth]{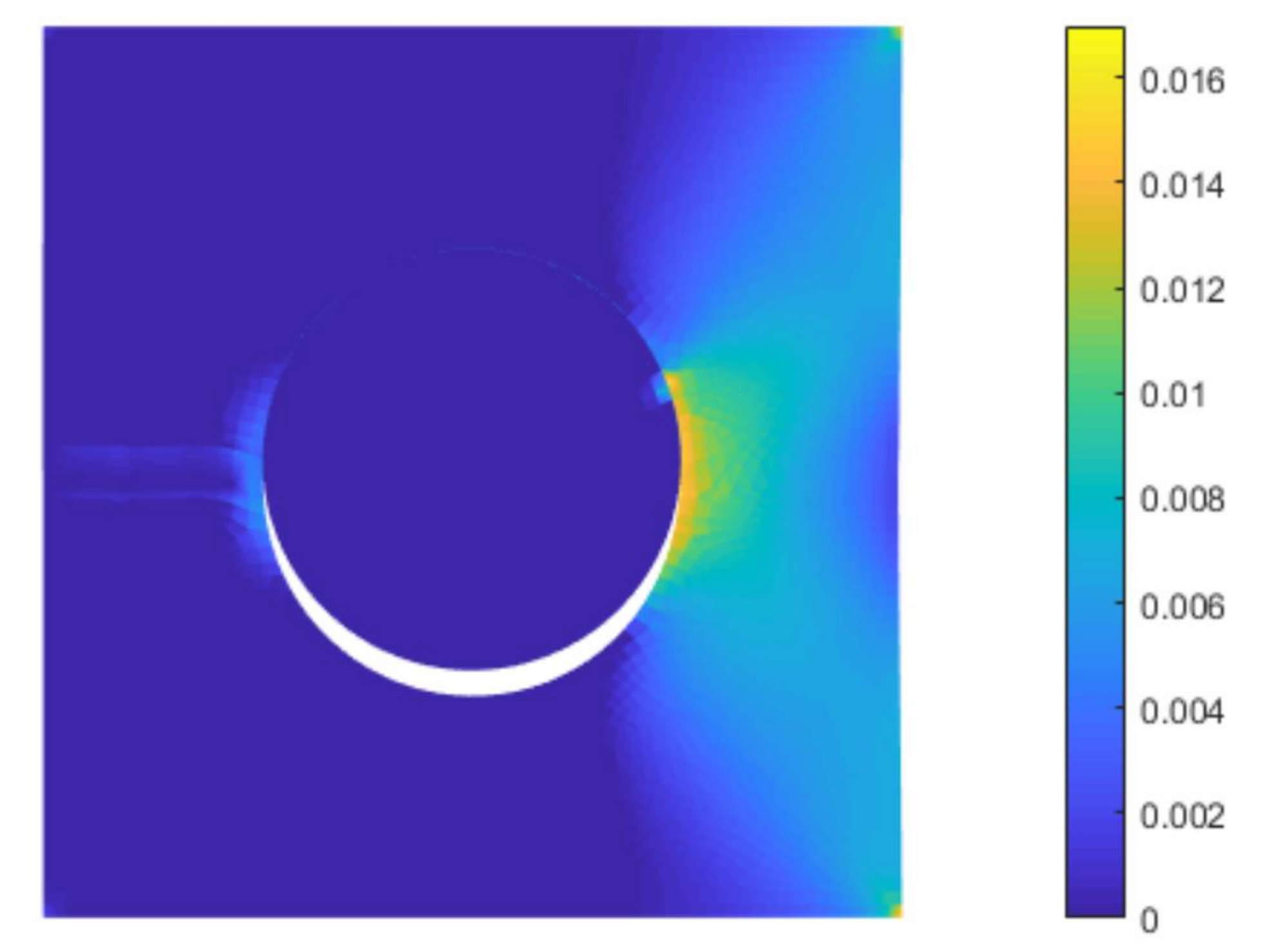}
	}
\caption{The effective plastic strain profiles in a composite with possible ductile matrix cracking and interface debonding for (a) stronger interface, and (b) weaker interface.}
\label{strong-weak-ductile-ep}
\end{figure}

The crack profiles after loading are shown in Fig.~\ref{strong-weak-ductile}. Similar to the results in Fig.~\ref{strong-weak}, for the weaker interface, matrix cracking and interface debonding both occur, and for the stronger interface, the matrix crack propagates along the interface while interface debonding is not triggered. In addition, for a stronger interface, the difference of the crack path between the brittle matrix cracking and the ductile matrix cracking shows the effect of plasticity. This is more explicitly shown in Fig.~\ref{strong-weak-ductile-ep}, which plots the corresponding effective plastic strain profiles.

\section{Conclusions}\label{conclusion}
In this work, we proposed a framework that combines the PFM and the CZM to investigate matrix cracking, interface debonding, and their competition in metal matrix fiber-reinforced composites in a minimalistic way. This approach enjoys the respective advantages of the CZM, effective in describing the evolution of the interface debonding, and the PFM, convenient in tracking the crack. The features of the framework are: (1) the interface debonding is not interfered by the phase field in the bulk, but governed merely by the CZM; (2) zero-thickness cohesive elements along the interface eliminate the need of regularizing the interface, giving rise to a cleaner model with fewer parameters; (3) there is hardly any restriction on the type of cohesive law to be used, unlike some existing works; (4) elastoplasticity of the matrix is taken into account and not affected by the interface debonding; (5) the competition of the two failure mechanisms, i.e., matrix cracking and interface debonding, is accurately captured. The proposed framework is verified using analytical solutions in the literature.

\section*{Acknowledgments}
This work is supported by the National Natural Science Foundation of China, grant No.~11972227 and the Natural Science Foundation of Shanghai, grant No.~19ZR1424200.

\bibliography{References}

\end{document}